\PassOptionsToPackage{switch,pagewise,columnwise,mathlines}{lineno}
\documentclass[fleqn,usenatbib]{mnras}

\usepackage{newtxtext,newtxmath} 
\usepackage[T1]{fontenc}
\usepackage{ae,aecompl}

\usepackage{etoolbox}
\usepackage[table,xcdraw]{xcolor}
\selectcolormodel{cmyk}
\usepackage{graphicx}  % Including figure files
\DeclareGraphicsExtensions{.pdf,.jpg,.png,.eps}
\graphicspath{{figs/}}
\usepackage{amsmath}   % Advanced maths commands
\usepackage{mathtools} % for DeclarePairedDelimiter
\usepackage{amsfonts}
\usepackage{booktabs}  % quality tables
\usepackage{multirow}
\usepackage{caption}
\usepackage{subcaption}
\usepackage{etoolbox}
\usepackage{siunitx}
\usepackage{comment}
\usepackage{xurl}

\newcommand{\nbpix}{N}
\newcommand{\nfourier}{K}
\newcommand{\nband}{L}

\newcommand{\nfacet}{Q}
\newcommand{\nwcoefs}{I}
\newcommand{\nblock}{B}

\newcommand{\ndata}{M}
\newcommand{\ncube}{C}
\newcommand{\ndict}{D}
\newcommand{\nreweight}{T}

\newcommand{\nreweightmax}{\nreweight}
\newcommand{\npdfb}{P}
\newcommand{\npdfbmin}{\npdfb_{\min}}
\newcommand{\npdfbmax}{\npdfb_{\max}}
\newcommand{\relvar}{\underline{\xi}}
\newcommand{\relvarpdfb}{\relvar_{\textnormal{\texttt{pdfb}}}}
\newcommand{\relvarrw}{\relvar_{\textnormal{\texttt{rw}}}}

\newcommand{\bs}[1]{\boldsymbol{#1}}

\DeclareMathOperator{\sgn}{sgn}

\newcommand{\minimize}[2]{\ensuremath{\underset{\substack{{#1}}}{\mathrm{minimize}}\;\;#2 }}

\DeclareMathOperator{\prox}{prox}

\DeclareMathOperator{\Diag}{Diag}

\newcommand{\argmind}[2]{\ensuremath{\underset{\substack{{#1}}}%
{\mathrm{argmin}}\;\;#2 }}

\DeclarePairedDelimiter{\norm}{\lVert}{\rVert}

\DeclareMathOperator{\SNR}{SNR}
\DeclareMathOperator{\aSNR}{aSNR}
\DeclareMathOperator{\sSNR}{sSNR}
\DeclareMathOperator{\iSNR}{iSNR}
\DeclareMathOperator{\runp}{\texttt{run}_\textsubscript{\texttt{pips}}}
\DeclareMathOperator{\run}{\texttt{run}}
\DeclareMathOperator{\cpup}{\texttt{cpu}_\textsubscript{\texttt{pips}}}
\DeclareMathOperator{\cpu}{\texttt{cpu}}

\definecolor{darkgreen}{cmyk}{0.8,0,0.8,0.45}
\definecolor{lightgreen}{cmyk}{0.8,0,0.8,0.25}
\usepackage[ruled,vlined,linesnumbered]{algorithm2e}

\SetCommentSty{mycommfont}
\SetKwInput{Parameter}{Parameters} % use \SetKwInOut{...}{...} if alignment between the entries is needed (align colons in the algorithm)
\let\oldnl\nl% store \nl in \oldnl
\newcommand{\nonl}{\renewcommand{\nl}{\let\nl\oldnl}} % remove line number for one line

\usepackage{tikz}
\usetikzlibrary{shapes,arrows,positioning,spy,quotes,arrows.meta}

\newcommand{\imdraw}[2]{
\begin{tikzpicture}[
    node distance = 0cm,
    inner sep = 0pt,
    node/.append style={ultra thin},
    ]
    \node[anchor=south west,inner sep=0] at (0,0) (image) {\includegraphics[keepaspectratio,width=#1]{#2}};
    \begin{scope}[x={(image.south east)},y={(image.north west)}]
        \draw[white] (0.48,0.22) rectangle ++(0.12,0.2);
        \draw[white] (0.46,0.75) rectangle ++(0.07,0.25);
        \draw[white] (0.23,0.75) rectangle ++(0.20,0.08);
    \end{scope}
\end{tikzpicture}
}

\usepackage[switch,pagewise,columnwise,mathlines]{lineno} 
\title[Faceted HyperSARA]{Parallel faceted imaging in radio interferometry via proximal splitting (Faceted HyperSARA): I. Algorithm and simulations}

\author[Thouvenin et al.]{Pierre-Antoine Thouvenin,$^{1,2}$%\thanks{The first three authors contributed equally to this work.}
Abdullah Abdulaziz,$^{1}$%\footnotemark[\value{footnote}]
Arwa Dabbech,$^{1}$
Audrey Repetti,$^{1,3}$
and Yves Wiaux$^{1}$\thanks{E-mail: y.wiaux@hw.ac.uk}
\\
$^{1}$Institute of Sensors, Signals and Systems, Heriot-Watt University, Edinburgh EH14 4AS, United Kingdom\\
$^{2}$Universit{\'e} de Lille, CNRS, Centrale Lille, UMR 9189 CRIStAL, F-59000 Lille, France\\
$^{3}$Department of Actuarial Mathematics \& Statistics, Heriot-Watt University, Edinburgh EH14 4AS, United Kingdom
}

\date{Accepted XXX. Received YYY; in original form ZZZ}

\pubyear{2021}

\begin{document}
\label{firstpage}
\pagerange{\pageref{firstpage}--\pageref{lastpage}}
\maketitle

% Abstract
\begin{abstract}
Upcoming radio interferometers are aiming to image the sky at new levels of resolution and sensitivity, with wide-band image cubes reaching close to the Petabyte scale for SKA. Modern proximal optimization algorithms have shown a potential to significantly outperform CLEAN thanks to their ability to inject complex image models to regularize the inverse problem for image formation from visibility data. They were also shown to be parallelizable over large data volumes thanks to a splitting functionality enabling the decomposition of the data into blocks, for parallel processing of block-specific data-fidelity terms involved in the objective function. Focusing on intensity imaging, the splitting functionality is further exploited in this work to decompose the image cube into spatio-spectral facets, and enable parallel processing of facet-specific regularization terms in the objective function, leading to the ``Faceted HyperSARA'' algorithm. Reliable heuristics enabling an automatic setting of the regularization parameters involved in the objective are also introduced, based on estimates of the noise level, transferred from the visibility domain to the domains where the regularization is applied. Simulation results based on a MATLAB implementation and involving synthetic image cubes and data close to Gigabyte size confirm that faceting can provide a major increase in parallelization capability when compared to the non-faceted approach (HyperSARA).
\end{abstract}

% Select between one and six entries from the list of approved keywords.
% Don't make up new ones.

%%%%%%%%%%%%%%%%%%%%%%%%%%%%%%%%%%%%%%%%%%%%%%%%%%

%%%%%%%%%%%%%%%%% BODY OF PAPER %%%%%%%%%%%%%%%%%%

\section{Introduction}\label{sec:intro}

Modern radio interferometers, such as the Karl G. Jansky Very Large Array (VLA)~\citep{Perley2011}, the LOw Frequency ARray (LOFAR)~\citep{Van2013} and the MeerKAT telescope~\citep{Jonas2018} generate extremely large volumes of data, with the aim of producing images of the radio sky at unprecedented resolution and dynamic range over thousands of spectral channels. The upcoming Square Kilometer Array (SKA) \citep{Dewdney2013} will form wide-band images about 0.56 Petabyte in size (around 1 Gigapixel per channel across tens of thousands of channels, and assuming double precision) from even larger visibility data volumes \citep{Scaife2020}. SKA is expected to bring answers to fundamental questions in astronomy, such as improving our understanding of cosmology and dark energy~\citep{Rawlings2004}, investigating the origin and evolution of cosmic magnetism~\citep{Gaensler2004} and probing the early universe where the first stars were formed~\citep{Carilli2004}. It is therefore of paramount importance to design efficient imaging algorithms which meet the capabilities of such powerful instruments. On the one hand, appropriate algorithms need to inject complex prior image models to regularize the inverse problem for image formation from visibility data, which only provide incomplete Fourier sampling. On the other hand, these algorithms need to be highly parallelizable in order to scale with the sheer amount of data and the large size of the wide-band image cubes to be recovered.

A plethora of radio-interferometric (RI) imaging approaches have been proposed in the literature, which can be classified into three main categories. A first class of methods is the celebrated CLEAN family~\citep[\emph{e.g.}][]{Hogbom1974,Schwab1983, Bhatnagar2004,Cornwell2008,Rau2011,Offringa2017}. In particular, \cite{Rau2011} proposed the multi-scale multi-frequency deconvolution algorithm (MS-MFS), leveraging Taylor series and multi-scale CLEAN to promote spectral smoothness of the wide-band image cube. More recently, \cite{Offringa2017} have proposed {a joint channel deconvolution CLEAN-based method}, where multi-scale CLEAN components are identified from the integrated residual image (\emph{i.e.} the sum of the residual images over all the channels). Albeit simple and computationally efficient, CLEAN-based algorithms are known to provide a limited imaging quality when reconstructing extended sources and emissions at high resolution and high sensitivity acquisition regimes. Restored CLEAN images tend to exhibit (i) artefacts due to the simplicity of the prior image models used, usually limited by the convolution of the model image with a ``CLEAN beam'' preventing any super-resolution beyond the spatial bandwidth of the observed data, and (ii) negative flux due to the addition of residuals into the final solution.

The second class of methods relies on Bayesian inference techniques~\citep[\emph{e.g.}][]{Sutton2006, Sutter2014, Junklewitz2015, Junklewitz2016, Arras2019}. 
For instance, \citet{Sutter2014} proposed a monochromatic Bayesian method based on a Markov chain Monte Carlo (MCMC) algorithm, considering a Gaussian image prior. Since MCMC sampling methods are computationally very expensive, an efficient variant was proposed by~\citet{Junklewitz2016, Arras2019} to perform approximate Bayesian inference, formulated in the framework of information theory. Importantly, Bayesian methods naturally enable the uncertainty about the image estimate to be quantified. However, this type of approaches cannot currently scale to the data regime expected from modern telescopes. 

The third class of approaches leverages optimization methods allowing sophisticated prior information to be considered, such as sparsity in an appropriate transform domain or spectral correlation~\citep[\emph{e.g.}][]{Wiaux2009,Li2011,Dabbech2012,Carrillo2012,Wenger2014,Garsden2015,Dabbech2015,Girard2015,Ferrari2015,Abdulaziz2016,Jiang2017,Abdulaziz2019}. From the perspective of optimization theory, the inverse imaging problem is addressed by defining an objective function, consisting of the sum of a data-fidelity term and a regularization term promoting a prior image model to compensate for the incompleteness of the visibility data. The sought image is estimated as a minimizer of this objective function, and is computed through iterative algorithms, which benefit from well-established convergence guarantees. For instance, \citet{Ferrari2015} promote spatial sparsity in a redundant wavelet domain and spectral sparsity \emph{via} a Discrete Cosine Transform. \citet{Wenger2014} promote spectra composed of a smooth contribution affected by local sparse deviations. In the last decade, Wiaux and collaborators proposed advanced image models: the \textit{average sparsity} prior in monochromatic imaging (SARA)~\citep{Carrillo2012,Carrillo2013,Carrillo2014,Onose2016eusipco,Onose2016,Onose2017,Pratley2017,Dabbech2018}, the \textit{low-rankness} and \textit{joint average sparsity} priors for
wide-band imaging (HyperSARA) \citep{Abdulaziz2016,Abdulaziz2017,Abdulaziz2019}, and the \textit{polarization constraint} for polarized imaging (Polarized SARA)~\citep{Birdi2018}\footnote{Associated software on the Puri-Psi webpage~\citep{puripsi}.}. These models have been reported to result in significant improvements in the reconstruction quality in comparison with state-of-the-art CLEAN-based imaging methods, at the expense of an increased computational cost~\citep{Onose2016, Birdi2018, Abdulaziz2019}.

Note that, from a Bayesian perspective, the objective function can be seen as the negative logarithm of a posterior distribution, with the minimizer corresponding to the Maximum \emph{A Posteriori} (MAP) estimate. Methods for uncertainty quantification by convex optimization have also been recently designed to assess the degree of confidence in specific structures appearing in the MAP estimate~\citep{Repetti2018,Repetti2019,Abdulaziz2019b}. In this work, we focus solely on image estimation.

Convex optimization offers intrinsically parallel algorithmic structures, such as proximal splitting methods~\citep{Combettes2011,Komodakis2015}. In such algorithmic structures, multi-term objective functions can be minimized, with the different terms handled in parallel at each iteration. Each term is involved through its so-called proximal operator, which acts as a simple denoising operator (\emph{e.g.} a sparsity regularization term will induce a soft-thresholding operator). The algorithms of the SARA family are all powered by an advanced proximal splitting method known as the primal-dual forward-backward (PDFB) algorithm~\citep{Condat2013, Vu2013,Pesquet2014}. The splitting functionality of PDFB is utilized in these approaches to enable the decomposition of data into blocks and parallel processing of the block-specific data-fidelity terms of the objective function, allowing the algorithm to be parallelized over large data volumes. The SARA family however models the image as a single variable, and the computational and memory requirements induced by complex regularization terms can be prohibitive for very large image sizes, in particular for wide-band imaging.

{Focusing on intensity imaging,} we address this bottleneck in the present work. We propose to decompose the target image cube into regular, content-agnostic, spatially overlapping spatio-spectral facets, with which are associated facet-specific regularization terms in the objective function. We further exploit the splitting functionality of PDFB to enable parallel processing of the regularization terms.

Note that faceting is not a novel paradigm in RI imaging; it has often been considered for calibration purposes in the context of wide-field imaging, assuming piece-wise constant direction-dependent effects. For instance, \citet{refId0} proposed an image tessellation scheme for LOFAR wide-field images, which has been leveraged by~\citet{Tasse2018} in the context of wide-field wide-band calibration and imaging. 
However, except for~\citet{Naghibzedeh2018} and~\citet{Tasse2018} (where the latter can accommodate any imaging algorithm), facet imaging has hitherto been mostly addressed with CLEAN-based algorithms to the best of the authors' knowledge. This class of approaches not only lacks theoretical convergence guarantees, but also does not offer much flexibility to accommodate advanced regularization terms. In contrast with~\citet{Naghibzedeh2018}, the proposed faceting approach does not need to be tailored to the content of the image, and thus enables the faceting to be used to balance computational costs across the available resources.

The remainder of the article is organized as follows. Section~\ref{sec:problem_statement} introduces the proposed faceted prior model and associated objective function underpinning Faceted HyperSARA. Heuristics are also proposed to set the regularization parameters involved in SARA, HyperSARA and Faceted HyperSARA, respectively. The associated algorithm is described in Section~\ref{sec:algorithm}, along with the different levels of parallelization exploited in the implementation considered. Performance validation is conducted on synthetic data in Section~\ref{sec:exp_simu}, in comparison with both HyperSARA and SARA. More precisely, the influence of spectral and spatial faceting is evaluated for a varying number of facets and spatial overlap, both in terms of reconstruction quality and computing time. 
Conclusions and perspectives are reported in Section~\ref{sec:conclusion}.

%% ==============================================================
%% PROBLEM STATEMENT
%% ==============================================================
 
\section{Prior model and objective function} \label{sec:problem_statement}

In this section focused on optimization-based approaches, the discrete version of the inverse problem for RI image formation from visibility data is first recalled. The general structure of the objective function is then formulated for the state-of-the-art SARA and HyperSARA approaches. Finally, the proposed spatio-spectral facets and the associated prior model are introduced, leading to the objective function considered in the proposed Faceted HyperSARA approach. New heuristics for automatic regularization parameter setting are also introduced for all three approaches. To ease the reading of the manuscript, repeatedly used notations are summarized in Table~\ref{tab:maths_notations}.

\begin{table}
    \centering
    \caption{Problem dimensions and mathematical notations}
    \label{tab:maths_notations}
    \begin{tabular}{ll} \toprule
        \multicolumn{2}{l}{\emph{Problem dimensions}} \\ \midrule
        $\nbpix$ & image size \\
        $\nfourier$ & size of the 2D oversampled Fourier space \\
        $\nblock$ & number of data blocks per channel \\
        $\nband$ & number of spectral channels \\
        $\ndata$ & number of visibilities \\
        $\ncube$ & number of spectral sub-cubes \\
        $\nfacet$ & number of spatial facets \\ \midrule
        \multicolumn{2}{l}{\emph{Mathematical notations}} \\ \midrule
        $\mathbf{I}_{\ndata \times \ndata}$ & identity matrix, of size $\ndata \times \ndata$ \\
        $\mathbf{0}_{\ndata}$ & vector of size $\ndata$ whose entries are all equal to 0\\
        $\mathbf{0}_{\ncube \times \nfacet}$ & matrix of size $\ncube \times \nfacet$ whose entries are all equal to 0\\
        $[\mathbf{Z}]_i$ & $i$th row of the matrix $\mathbf{Z}$ \\
        $[\mathbf{Z}]_{i,j}$ & element in position $(i,j)$ of the matrix $\mathbf{Z}$ \\
        $\mathbf{Z}^\dagger$ & adjoint (conjugate transpose) of the matrix $\mathbf{Z}$ \\
        $\sgn$ & element-wise sign function \\
        $\iota_C$ & indicator function of the set $C$ \\
        $\vert \cdot \vert$ & element-wise absolute value \\
        $\Vert \cdot \Vert_{\text{S}}$ & spectral norm of a linear operator \\
        $\Vert \cdot \Vert_{*}$ & nuclear norm \\
        $\Vert \cdot \Vert_{2,1}$ & $\ell_{2,1}$ norm \\
        $\Vert \cdot \Vert_{*,\overline{\bs{\omega}}}$ & weighted nuclear norm, with weights $\overline{\bs{\omega}}$ (see~\eqref{eq:weighted_nuclear_norm}) \\
        $\Vert \cdot \Vert_{2,1,\bs{\omega}}$ & weighted $\ell_{2,1}$ norm, with weights $\bs{\omega}$ (see~\eqref{eq:weighted_l21_norm}) \\
        \bottomrule
    \end{tabular}
\end{table}

%--------------------------------------------
\subsection{Wide-band inverse problem}

Wide-band RI imaging consists in estimating unknown radio images over $\nband$ frequency channels. Focusing on intensity imaging and assuming a small field of view on the celestial sphere, at a given frequency, each pair of antennas acquires a noisy Fourier component of the sky surface brightness, called a visibility. The associated Fourier mode (also called $uv$-point) is given by the projection of the corresponding baseline, expressed in units of the observation wavelength, {onto} the plane perpendicular to the line of sight~\citep{Thompson2007}. The collection of the sensed Fourier modes, accumulated from the baselines over the whole duration of the observation provides an incomplete coverage of the 2D Fourier plane (also called $uv$-plane) of the image of interest. The RI measurements can be modeled for each frequency channel indexed by $l \in \{1, \dotsc, \nband \}$ as~\citep{Abdulaziz2016,Abdulaziz2019}
\begin{linenomath}
\begin{equation} \label{eq:model}
 \mathbf{y}_l = \bs{\Phi}_l \overline{\mathbf{x}}_l + \mathbf{n}_l, \: \text{with } \bs{\Phi}_l = \mathbf{G}_l \mathbf{FZ}.
\end{equation}
\end{linenomath}
In this equation, $\mathbf{y}_l \in \mathbb{C}^{\ndata_l}$ is the vector of $\ndata_l$ visibilities acquired in the channel $l~\in~\{ 1, \dotsc, \nband \}$. 
The vector $\overline{\mathbf{x}}_l \in \mathbb{R}_+^\nbpix$ is the underlying image, with $\nbpix$ the number of image pixels. The vector $\mathbf{n}_l \in \mathbb{C}^{\ndata_l}$ represents the measurement noise modeled as a realization of a complex white\footnote{This model assumes noise with constant variance across visibilities within each channel. In full generality the noise is not white, in which case a diagonal whitening matrix $\bs{\Theta}_l \in \mathbb{R}^{\ndata_l \times \ndata_l}$ with entries equal to the inverse noise standard deviation per visibility can be applied to the original measurement vector, ensuring that the resulting visibility vector is indeed affected by white noise. In this case problem~\eqref{eq:model} still holds, with a measurement operator $\bs{\Phi}_l = \bs{\Theta}_l\mathbf{G}_l\mathbf{FZ}$, and $\mathbf{y}_l$ and $\mathbf{n}_l$ the visibility and noise vectors after application of the whitening matrix. This is known as natural weighting, which, among other weighting schemes in the field, is preferred by the algorithms of the SARA family to ensure the negative log-likelihood interpretation of the data-fidelity terms leveraging a simple $\ell_2$ norm of the residual visibilities~\citep{Wiaux2009,Onose2017}.} Gaussian noise, \emph{i.e.}, a complex Gaussian distribution $\mathcal{CN}(\mathbf{0}_{\ndata_l}, \sigma^2_l \mathbf{I}_{\ndata_l \times \ndata_l})$ with mean $\mathbf{0}_{\ndata_l}$ and covariance matrix $\sigma^2_l \mathbf{I}_{\ndata_l \times \ndata_l}$. The measurement operator $\bs{\Phi}_l$ is also composed of a zero-padding and scaling operator $\mathbf{Z} \in \mathbb{R}^{\nfourier \times \nbpix}$ (with $\nfourier$ the number of points after zero-padding), the 2D Discrete Fourier Transform represented by the matrix $\mathbf{F} \in \mathbb{C}^{\nfourier \times \nfourier}$, and a sparse non-uniform Fourier transform interpolation (or de-gridding) matrix $\mathbf{G}_l \in \mathbb{C}^{\ndata_l \times \nfourier}$. Each row of $\mathbf{G}_l$ contains a compact support interpolation (or de-gridding) kernel of size $S$ centered at the corresponding $uv$-point~\citep{Fessler2003}, enabling the computation of the associated visibility from the neighbouring discrete Fourier points. 

Note that at the sensitivity of the modern radio telescopes, direction-dependent effects (DDEs), which can be of atmospheric or instrumental origin, complicate the RI measurement equation. More precisely, for each visibility, the sky surface brightness is modulated by the product of the DDE patterns specific to each antenna. The DDEs are typically unknown and need to be calibrated jointly with the imaging process~\citep{Repetti2017,Repetti2017spie,Thouvenin2018b,Birdi2019,Dabbech2021}. One exception to this, is the so-called $w$-term induced by the projection of the baseline associated with each visibility on the line of sight, which becomes non-negligible on wide fields of view, and thus acts as a known DDE. Correcting for this DDE is, however, computationally demanding.
When focusing on imaging only, all DDEs are assumed to be either known or properly estimated by a prior calibration procedure, and can be easily integrated into the forward model \eqref{eq:model} by building extended de-gridding kernels into each row of $\mathbf{G}_l$, which are the outcome of the convolution of the non-uniform Fourier transform kernel with a compact-support representation of the Fourier transform of the involved DDEs. While the typical size $S$ of the non-uniform Fourier transform kernels is $7 \times 7$~\citep{Fessler2003}, this can increase significantly in the presence of DDEs. For example, for real VLA observations of Cyg~A, \citet{Dabbech2021} have considered calibration kernels in the spatial Fourier domain of size $5 \times 5$ for each antenna, to correct for pointing errors. Following a convolution with a $7 \times 7$ non-uniform Fourier transform kernel in the form of a Kaiser-Bessel kernel, the resulting  de-gridding kernel is of size $S = 15 \times 15$. However, on wide fields of view, the additional $w$-term can further increase the size by orders of magnitude. Nevertheless, provided that the extended de-gridding kernel size remains well below the spatial bandwidth of interest \citep{wiaux09,Dabbech2017}, the overall de-gridding matrix remains sparse.

\begin{figure}
    \centering
    \hspace{0.3cm} (a) Full data cube \hfill \hspace{0.65cm} (b) Per channel data \hfill (c) Data blocks \hspace{0.2cm} \hfill\\
    \vspace{0.1cm}
    \includegraphics[keepaspectratio,width=0.45\textwidth]{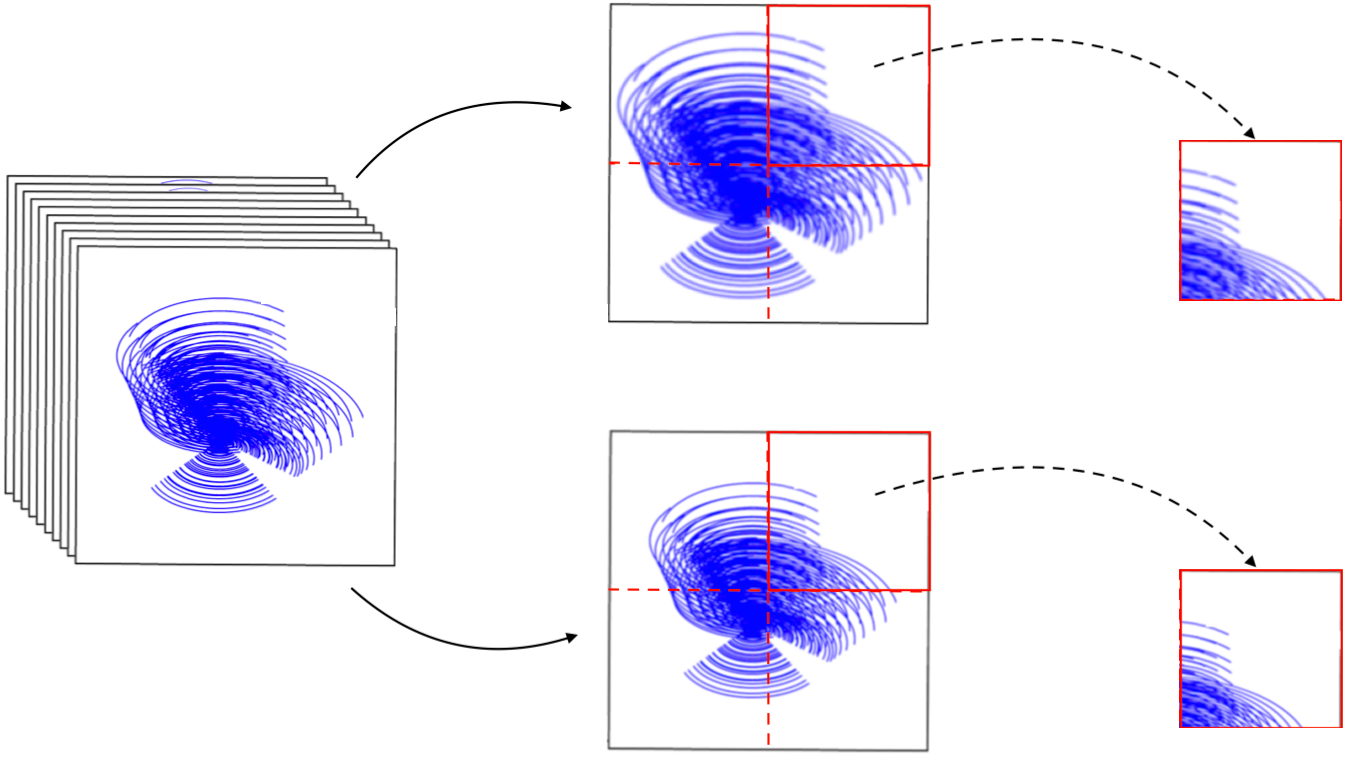}
    \caption{Illustration of the data blocking strategy. Starting from the full data cube (a), each of the $\nband$ channels represented in (b) is decomposed into $\nblock$ data blocks (c). The data blocks for each channel are associated with separate data-fidelity terms in the objective function, processed by independent workers.}
    \label{fig:blocking}
\end{figure}

When addressing the model~\eqref{eq:model}, a first bottleneck arises from the sheer volume of the data. To overcome it, \citet{Onose2016,Onose2017} have proposed a data blocking strategy, which has been exploited in the context of wide-band imaging by \citet{Abdulaziz2019}. The visibility vectors $\mathbf{y}_l$ are decomposed into $\nblock$ blocks, $\mathbf{y}_l = (\mathbf{y}_{l,b})_{1 \leq b \leq \nblock}$, which can be handled in parallel by advanced imaging algorithms, with controlled memory and computational cost per block. The data model~\eqref{eq:model} can be formulated for any $(l,b) \in \{1, \dots, \nband\} \times \{1, \dotsc, \nblock\}$ as
\begin{linenomath}
\begin{equation} \label{eq:model_with_blocking}
 \mathbf{y}_{l,b} = \bs{\Phi}_{l,b} \overline{\mathbf{x}}_l + \mathbf{n}_{l,b}, \: \text{with } \bs{\Phi}_{l,b} = \mathbf{G}_{l,b} \mathbf{FZ},
\end{equation}
\end{linenomath}
where $\mathbf{y}_{l,b} \in \mathbb{C}^{\ndata_{l,b}}$ is the vector of $\ndata_{l,b}$ visibilities associated with the $b$-th block in the channel $l$, $\bs{\Phi}_{l} = (\bs{\Phi}_{l,b})_{1 \leq b \leq \nblock}$, and $\mathbf{n}_l = (\mathbf{n}_{l,b})_{1 \leq b \leq \nblock}$. Different blocking strategies can be adopted, \emph{e.g.} based on a tessellation of the $uv$-space into balanced sets of visibilities~\citep{Onose2016}, or on a decomposition of the data into groups of time slots~\citep{Dabbech2018} (see Figure~\ref{fig:blocking}).

%--------------------------------------------
\subsection{General form of the objective function}

Estimating the underlying wide-band sky image $\overline{\mathbf{X}} = (\overline{\mathbf{x}}_l)_{1 \leq l \leq \nband}$ from incomplete Fourier measurements is a severely ill-posed inverse problem, which calls for powerful regularization terms to encode a prior image model. In this context, wide-band RI imaging can be formulated as the following constrained optimization problem
\begin{linenomath}
\begin{equation} \label{eq:problem}
	\minimize{\mathbf{X} = (\mathbf{x}_l)_{1 \leq l \leq \nband} \in \mathbb{R}_+^{\nbpix \times \nband}} \sum_{l =1}^\nband \sum_{b =1}^\nblock \iota_{\mathcal{B}(\mathbf{y}_{l,b}, \varepsilon_{l,b})} \bigl( \bs{\Phi}_{l,b}\mathbf{x}_{l} \bigr) + r(\mathbf{X}),
\end{equation}
\end{linenomath}
where the indices $(l,b) \in \{ 1, \dotsc , \nband \} \times \{ 1, \dotsc , \nblock \}$ refer to a data block $b$ of channel $l$, $\mathcal{B}(\mathbf{y}_{l,b},\varepsilon_{l,b}) = \bigl \{ \mathbf{z} \in \mathbb{C}^{\ndata_{l,b}} \mid \norm{\mathbf{z} - \mathbf{y}_{l,b}}_2 \leq \varepsilon_{l,b} \bigr\}$ denotes the $\ell_2$-ball centred in $\mathbf{y}_{l,b}$ of radius $\varepsilon_{l,b} > 0$. The parameter $\varepsilon_{l,b}^2 = \varepsilon_{l}^2 \ndata_{l,b} / \ndata_l$, with $\varepsilon_{l}^2 = \sigma^2_l (\ndata_{l} + 2 \sqrt{\ndata_{l}})$, reflects the noise statistics~\citep[see][for further details]{Onose2016,Carrillo2012}. The notation $\iota_{\mathcal{B}(\mathbf{y}_{l,b},\varepsilon_{l,b})}$ denotes the indicator function of the $\ell_2$ ball $\mathcal{B}(\mathbf{y}_{l,b}, \varepsilon_{l,b})$. Specifically, let $\mathcal{C}$ be a non-empty, closed, convex subset of $\mathbb{C}^\nbpix$, then $\iota_{\mathcal{C}}$ denotes the indicator function of $\mathcal{C}$, defined by $\iota_{\mathcal{C}} (\mathbf{z}) = +\infty$ if $\mathbf{z} \in \mathcal{C}$ and 0 otherwise. On the one hand, the indicator functions $\iota_{\mathcal{B}(\mathbf{y}_{l,b}, \varepsilon_{l,b})}$ are utilized in the data-fidelity terms to ensure the consistency of the modeled data with the measurements and reflect the white Gaussian nature of the noise~\citep{Carrillo2012}. On the other hand, the function $r$ encodes a prior model of the unknown image cube. The priors characterizing the state-of-the-art SARA and HyperSARA approaches, as well as the proposed Faceted HyperSARA approach, are discussed in what follows. On a final note, an additional non-negativity prior is imposed in all approaches of the SARA family when focusing on intensity imaging. The prior generalizes to the polarization constraint when solving for the Stokes parameters~\citep{Birdi2018, Birdi2019}).

%--------------------------------------------
\subsection{State-of-the-art average sparsity priors} \label{ssec:priors}

Sparsity-based priors combined with optimization techniques have proved to be a powerful regularization for astronomical imaging, in particular in the context of radio interferometry \citep{Wiaux2009,Carrillo2012,Garsden2015,Onose2016}. 
These techniques aim to solve the underlying image recovery problem by enforcing sparsity of the estimated image in an appropriate domain. 

In this context, the prior of choice is the $\ell_0$ pseudo-norm, which counts the number of non-zero coefficients of its argument \citep{Donoho2006}. However, minimizing this function, which is neither convex nor smooth, is a NP-hard problem. A common alternative consists in replacing it by its convex envelope, the $\ell_1$ norm \citep{Donoho1989,Donoho1992}. When combined with other convex terms, \emph{e.g.} the non-negativity constraint and the $\ell_2$-ball data-fidelity constraint in~\eqref{eq:problem}, $\ell_1$-norm priors form a convex objective function, which can be efficiently minimized by iterative algorithms under well-established guarantees on the convergence of the iterates towards a global minimum.

Although the $\ell_1$ prior has been widely used over the last decades to promote sparsity, it induces an undesirable dependence on the coefficients' magnitude. Indeed, unlike the $\ell_0$ pseudo-norm, the $\ell_1$ norm disproportionally penalizes the larger coefficients. To address this imbalance, a log-sum prior can be used. In particular,
\citet{Candes2008, Candes2009} proposed to use a majorization-minimization framework~\citep{Hunter2004} to minimize objective functions with a log-sum prior. This leads to a reweighted-$\ell_1$ approach, which consists in minimizing a sequence of convex objectives involving a weighted $\ell_1$ norm prior, acting as convex relaxations of the log-sum prior. In practice, sequentially minimizing convex problems with weighted-$\ell_1$ priors is indeed much simpler than minimizing a non-convex problem involving a log-sum prior. From a convergence point of view, multiple works have recently shown that the solution provided by a reweighted-$\ell_1$ procedure corresponds to a critical point of the log-sum function~\citep{Ochs2015, Geiping2018, Ochs2019, Repetti2019b}. In the following paragraphs, the SARA~\citep{Carrillo2012} and HyperSARA~\citep{Abdulaziz2019} log-sum priors are presented, and will be considered as benchmarks to assess the proposed spatio-spectral faceted prior. 

%--------------------------------------------
\subsubsection{SARA prior}
\label{Ssec:SARA}

The image prior underpinning the Sparse Averaging Reweighted Analysis (SARA) has proved powerful for RI imaging~\citep{Carrillo2012, Onose2016, Abdulaziz2019}. It promotes sparsity by minimizing a log-sum prior, considering a highly redundant transformed domain $\bs{\Psi}^\dagger \in \mathbb{R}^{I \times \nbpix}$, where $\nbpix$ is the number of image pixels, and $\nwcoefs$ is the number of coefficients obtained after applying the linear transform $\bs{\Psi}^\dagger$. The dictionary $\bs{\Psi}^\dagger$ is defined as the concatenation of $\ndict = 9$ orthonormal bases (first eight Daubechies wavelets and the Dirac basis), re-normalized by $\sqrt{D}$ to ensure that $\bs{\Psi}\bs{\Psi}^\dagger = \mathbf{I}_{\nbpix \times \nbpix}$. The log-sum prior addressed by SARA, which promotes average sparsity over the selected wavelet bases, is of the form
\begin{linenomath}
\begin{equation} \label{eq:log-sum-gen}
 r(\mathbf{X}) = \sum_{l=1}^\nband \widetilde{\mu}_l \sum_{i=1}^I \widetilde{\upsilon}_l \log\bigg( \frac{| [ \bs{\Psi}^\dagger \mathbf{X}]_{i,l} |}{\widetilde{\upsilon}_l} + 1 \bigg),
\end{equation}
\end{linenomath}
where $\widetilde{\mu}_l>0$ and $\widetilde{\upsilon}_l>0$ are regularization parameters, and $[ \bs{\Psi}^\dagger \mathbf{X}]_{i,l}$ denotes the $(i,l)$-th coefficient of $\bs{\Psi}^\dagger \mathbf{X}$. 

This prior is fully separable with respect to the spectral channels, similarly to the data-fidelity term in~\eqref{eq:problem}. In this setting, the wide-band objective function naturally separates into $\nband$ independent, single-channel objective functions underpinning the monochromatic SARA approach defined by~\citet{Carrillo2012,Onose2016}. This approach is therefore highly parallelizable, and will be taken as a reference in terms of computing time. Each single-channel term in the objective~\eqref{eq:problem} is independently minimized with a reweighting approach leveraging PDFB. In particular, the splitting functionality of this proximal algorithm enables the block-specific data-fidelity terms to be handled in parallel. In theory, the parameters $\widetilde{\mu}_l$ do not affect the minimizers of the objective function, as~\eqref{eq:problem}-\eqref{eq:log-sum-gen} can be reformulated as a log-sum minimization problem with constraints (for which rescaling the function $r$ amounts to rescaling to full objective function, with no effect on its minimizers). Note however that the convergence guarantees of the proposed algorithm are asymptotic, and that the objective function is not convex. These considerations imply that the regularization parameter has a practical impact on the observed convergence speed of the algorithm, and on the critical point numerically obtained at convergence.

\paragraph*{A new heuristic for automatic regularization parameter setting with SARA.} To set the regularisation parameters involved in \eqref{eq:log-sum-gen}, \citet{Onose2016} suggested to fix $\widetilde{\mu}_l \in [10^{-5}, 10^{-3}]$. As for the parameter $\widetilde{\upsilon}_l$, used to avoid reaching zero values in the argument of the log terms, \citet{Carrillo2012} proposed to set it to an estimate of the standard deviation of the noise in the wavelet domain.

Building on these empirical considerations, we propose a new heuristic to automatically set the regularization parameters of the SARA objective function (which will also generalize to HyperSARA and Faceted HyperSARA). On the one hand, $\widetilde{\upsilon}_l$ is understood as a floor level for the wavelet coefficients. On the other hand, $\widetilde{\mu}_l$ explicitly acts as a soft-thresholding parameter \citep[Algorithm 2]{Onose2016}, setting to zero negligible values, and can therefore also be understood as a floor level for the wavelet coefficients. The floor level value needed for both $\widetilde{\upsilon}_l$ and $\widetilde{\mu}_l$ can naturally be set to an estimate of the standard deviation of the noise obtained by propagating the measurement noise from the visibility domain, successively to the image and the wavelet domains.

In order to compute the standard deviation of the noise in the image and wavelet domains, we consider a simple ``noise uncorrelation'' approximation consisting in assuming that i.i.d.~Gaussian noise of standard deviation $\sigma$ remains i.i.d.~Gaussian under a linear transform $\mathbf{A}$ (inverse measurement operator or sparsity dictionary operator), with a standard deviation rescaled by the spectral norm of $\mathbf{A}$, denoted by $\norm{\mathbf{A}}_\text{S}$\footnote{In practice, the spectral norm $\norm{\mathbf{A}}_\text{S} = \sqrt{\norm{\mathbf{A}^\dagger \mathbf{A}}_\text{S}}$ can be computed via the power method~(see \emph{e.g.}~\citet
%[Section 7.3.1]
{Golub2013}), which exclusively involves applications of the operator $\mathbf{A}^\dagger\mathbf{A}$. Numerical convergence of the power method (considering a relative variation of the solution lower than $10^{-4}$) has been systematically observed within a maximum of 100 iterations for all the examples considered in this paper.)}. This amounts to approximating the covariance matrix after linear transform, $\sigma^2\mathbf{A}\mathbf{A}^\dagger$, with $\sigma^2\norm{\mathbf{A}}^2_\text{S}\mathbf{I}$, preserving its norm while discarding the correlation structure. 
Firstly, to propagate the noise in the image domain, we consider the following normalized version of the inverse problem \eqref{eq:model}:  $\mathbf{y}_l/\norm{\bs{\Phi}_l}_\text{S} = \bs{\Phi}_l \mathbf{x}_l /\norm{\bs{\Phi}_l}_\text{S} + \mathbf{n}_l/\norm{\bs{\Phi}_l}_\text{S}$, where the measurement operator $\bs{\Phi}_l /\norm{\bs{\Phi}_l}_\text{S}$ has unit spectral norm, and $\mathbf{n}_l/\norm{\bs{\Phi}_l}_\text{S}$ is a rescaled i.i.d.~Gaussian noise of standard deviation %
\begin{equation}\label{normalisednoise}
\overline{\sigma}_l=\frac{\sigma_l}{\norm{\bs{\Phi}_l}_\text{S}}.
\end{equation}
We argue that this is the proper normalization as, when applying the adjoint of the normalized measurement operator to transfer the inverse problem to the image domain, the problem reads $\bs{\Phi}_l^\dagger \mathbf{y}_l/\norm{\bs{\Phi}_l}^2_\text{S} = \bs{\Phi}_l^\dagger\bs{\Phi}_l \mathbf{x}_l /\norm{\bs{\Phi}_l}^2_\text{S} + \bs{\Phi}_l^\dagger \mathbf{n}_l/\norm{\bs{\Phi}_l}^2_\text{S}$, where $\bs{\Phi}_l^\dagger\bs{\Phi}_l/\norm{\bs{\Phi}_l}^2_\text{S}$ represents the convolution by a properly normalized dirty beam, ensuring that the dirty image is at the same scale as the original image. In this context, the noise in the image domain is obtained by application of a linear operator with unit spectral norm ($\mathbf{A} \rightarrow \bs{\Phi}_l ^\dagger /\norm{\bs{\Phi}_l}_\text{S}$) so that, under the uncorrelation approximation, the standard deviation $\overline{\sigma}_l$ of the noise is preserved when transferred from data to image domain. 
Secondly, the noise in the wavelet domain results from applying $\bs{\Psi}^\dagger$ to the image domain noise (see the prior \eqref{eq:log-sum-gen}). Since the SARA dictionary also has unit spectral norm, and applying once more the uncorrelation approximation, the estimate of the noise level in the wavelet domain is the same as in the image domain: $\overline{\sigma}_l$.
The proposed heuristic thus reads 
\begin{linenomath} \label{eq:heuristic_sara}
\begin{equation}
    \widetilde{\mu}_l = \widetilde{\upsilon}_l =\overline{\sigma}_l.
\end{equation}
\end{linenomath}

%--------------------------------------------
\subsubsection{HyperSARA prior}
\label{Ssec:hyperSARA}

Compared to the SARA prior, the HyperSARA prior~\citep{Abdulaziz2019} aims to explicitly promote spectral correlations. In particular, it promotes low-rankness of the wide-band image resulting from the correlation of its channels, as well as average spatial sparsity over all the frequency channels. The log-sum prior of interest is of the form
\begin{linenomath}
\begin{multline}
    r (\mathbf{X}) = \overline{\mu} \sum_{j=1}^J \overline{\upsilon} \log \bigg( \frac{| s_j(\mathbf{X}) |}{\overline{\upsilon}} + 1 \bigg) \\
    + \mu \sum_{i=1}^I \upsilon \log \bigg( \frac{\| [ \bs{\Psi}^\dagger \mathbf{X} ]_i \|_2}{\upsilon} + 1 \bigg),
    \label{eq:hypersara0}
\end{multline}
\end{linenomath}
where $(\overline{\mu},  \overline{\upsilon},\mu, \upsilon) \in ]0, +\infty[^4$ are regularization parameters, $[ \bs{\Psi}^\dagger \mathbf{X} ]_i$ denotes the $i$-th row of $\bs{\Psi}^\dagger \mathbf{X}$ and $J = \min\{\nbpix, \nband\}$. The vector $\big( s_j (\mathbf{X}) \big)_{1 \le j \le J}$ contains the singular values of $\mathbf{X}$ obtained by a singular value decomposition (SVD).

HyperSARA has been shown to produce image cubes with superior quality when compared to SARA and the wide-band CLEAN-based approach WSClean \citep{Abdulaziz2019}. The main reason for this is that the number of degrees of freedom to be reconstructed when adding frequency channels increases more slowly than the amount of data due to the correlation between the channels. Moreover, since the magnitude of the spatial frequency sensed by an antenna pair is proportional to the observation frequency, the $uv$-coverage at a higher frequency channel is a dilated version of the $uv$-coverage at a lower frequency, with the dilation parameter between two channels given by the frequency ratio. Consequently, the data at higher frequency channels provide higher spatial frequency information for the lower frequency channels, leading to a more accurate image reconstruction process in terms of both resolution and dynamic range. HyperSARA will thus be taken as a reference in terms of imaging quality in Section~\ref{sec:exp_simu}.

In practice, the wide-band imaging problem~\eqref{eq:problem} is also solved with a reweighting procedure relying on PDFB, in which the splitting functionality allows the block-specific data-fidelity terms to be processed in parallel. The regularization terms at the core of HyperSARA are however not separable, with the full image cube modeled as a single variable. This entails memory and computing requirements scaling with the size of the full image cube. The focus of the present contribution is to address this bottleneck by introducing spatio-spectral image facets. 

In theory, only the ratio between $\overline{\mu}$ and $\mu$ affects the minimizers of the objective function~\eqref{eq:problem}. A similar remark as in Section 2.3.2 can however be made about the practical influence of $\overline{\mu}$ and $\mu$ in the algorithm, which converges to a critical point of \eqref{eq:problem}. 

\paragraph*{A new heuristic for automatic regularization parameter setting with HyperSARA.}
We generalize the SARA heuristic, building from the same observation that $\mu$ and $\overline{\mu}$ respectively act as soft-thresholding parameters in the wavelet and singular value domains respectively \citep{Abdulaziz2019}, to be set to corresponding noise level values, similarly to $\upsilon$ and $\overline{\upsilon}$.

With regards to the wavelet domain, the noise resulting from application of $\bs{\Psi}^\dagger$, approximately i.i.d.~Gaussian with standard deviation $\overline{\sigma}_l$, is further transformed by the computation of an $\ell_2$ norm across frequencies to reach the domain of application of $\upsilon$ and $\mu$ (see the prior \eqref{eq:hypersara0}). We further assume for simplicity that the standard deviation of each i.i.d.~Gaussian component before taking the $\ell_2$ norm is constant across frequencies and approximate it to the square root of the average variance $\overline{\sigma}^2 = 1/\nband \sum_{l=1}^\nband \overline{\sigma}^2_l$.

The $\ell_2$ norm of the noise consequently follows a simple Chi distribution re-scaled by $\overline{\sigma}$. The heuristic for $\mu$ and $\upsilon$ results from approximating the corresponding noise level to one standard deviation above the mean of that Chi distribution:
\begin{linenomath}
\begin{equation} \label{eq:heuristic_hypersara_sparsity}
    \mu = \upsilon = \overline{\sigma}[ m^{(\chi_{\nband})}+s^{(\chi_{\nband})}],
\end{equation}
\end{linenomath}
where $m^{(\chi_{\nband})}$ and $s^{(\chi_{\nband})}$ identify the mean and standard deviation of a Chi distribution with $\nband$ degrees of freedom, denoted $\chi_{\nband}$. Note that when $\nband$ is large, the concentration of measure phenomenon implies that the $\chi_{\nband}$ distribution is concentrated around its mean, leading to $[ m^{(\chi_{\nband})}+s^{(\chi_{\nband})}]\simeq\sqrt{\nband}$, and equation \eqref{eq:heuristic_hypersara_sparsity} naturally boils down to estimating the noise level as the $\ell_2$ norm $\sqrt{\nband}\overline{\sigma}$ of the vector of standard deviations across channels.

With regards to the singular value domain, the noise is interpreted as resulting from the application of a unitary (SVD) transform to the image domain noise, which we assume preserves the Frobenius norm $\sqrt{\nbpix\nband}\overline{\sigma}$ of the matrix of the standard deviations. Acknowledging that the noise diagonalizes from $\nbpix\nband$ down to $J$ values in the SVD process leads to the following estimate of the noise standard deviation and the heuristic for $\overline{\mu}$ and $\overline{\upsilon}$:
\begin{linenomath}
\begin{equation} \label{eq:heuristic_hypersara_low_rank}
    \overline{\mu} = \overline{\upsilon} = \overline{\sigma} \sqrt{\frac{\nbpix\nband}{J}}.
\end{equation}
\end{linenomath}

\begin{figure}
	\centering
	(a) Full image cube \hspace{0.4cm} (b) Spectral sub-cubes \hspace{0.15cm} (c) Facets \& weights\\
	\vspace{0.1cm}
	\includegraphics[keepaspectratio,width=0.45\textwidth]{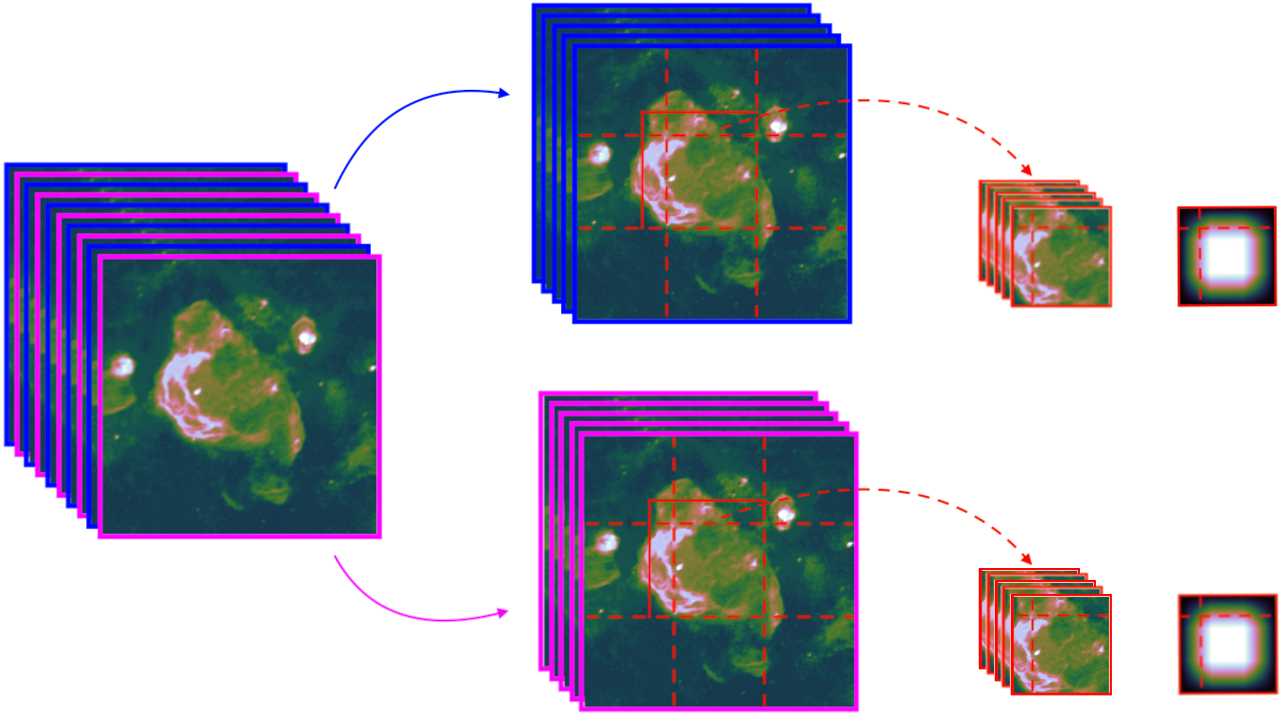}
    \caption{Illustration of the proposed faceting scheme, using a 2-fold spectral interleaving process and 9-fold spatial tiling process. The full image cube variable (a) is divided into two spectral sub-cubes (b) with interleaved channels (for a 2-fold interleaving, even and odd channels respectively define a sub-cube). Each sub-cube is spatially faceted. A regular tessellation (dashed red lines) is used to define spatio-spectral tiles. The spatio-spectral facets result from the augmentation of each tile to produce an overlap between facets (solid red lines). Panel (c) shows a single facet (left), as well as the spatial weighting scheme (right) with linearly decreasing weights affecting pixels involved in multiple contiguous facets. Note that, though the same tiling process is underpinning the low-rankness and the average joint sparsity regularization terms, the definition of the appropriate overlap region is specific to each of these terms (\emph{via} the selection operators $\widetilde{\mathbf{S}}_q$ and $\mathbf{S}_q$ in~\eqref{eq:facet_hypersara}).}
    \label{fig:facets}
\end{figure}

%--------------------------------------------
\subsection{Faceting and Faceted HyperSARA prior} \label{sec:faceting_prior}

The proposed Faceted HyperSARA prior builds on the HyperSARA prior, distributing both the average sparsity and the low-rankness prior over multiple spatio-spectral facets to split the computing and memory costs of HyperSARA over a larger number of cores. We propose to decompose the 3D image cube into $\nfacet \times \ncube$ spatio-spectral facets as illustrated in Figure~\ref{fig:facets} and detailed in what follows.

\subsubsection{Spectral faceting} \label{sec:spectral_faceting}
The wide-band image cube can first be decomposed into separate image sub-cubes composed of a subset of the frequency channels, with a separate prior taken for each sub-cube. Since the data-fidelity terms are channel-specific, the overall objective function~\eqref{eq:problem} reduces to the sum of independent objective functions for each sub-cube. These smaller-size wide-band imaging sub-problems (smaller data sets, and smaller image volumes) can thus be solved independently in parallel.
Taken to the extreme, this simple spectral faceting can be used to separate all channels and proceed with single-channel reconstructions (corresponding to SARA), yet, however loosing completely the advantage of leveraging the correlations between frames to improve image precision. In practice, it is essential to keep an appropriate number of frames per sub-cube in order to take advantage of this correlation. In addition, given that data at higher frequency channels provide higher spatial frequency information for the lower frequency channels, it is of critical importance that the whole extent of the frequency band of observation is exploited in each channel reconstruction. In this context, we propose to decompose the cube into channel-interleaved spectral sub-cubes, each of which results from a uniform sub-sampling of the whole frequency band (see Figure~\ref{fig:facets} (b)). 
We thus decompose the original inverse problem~\eqref{eq:model} into $\ncube$ independent, channel-interleaved sub-problems, each considering $\nband_c$ channels from the original data cube, with $\nband= \nband_1 + \ldots + \nband_\ncube$. For each sub-cube indexed by $c\in \{1, \ldots, \ncube\}$, the data model~\eqref{eq:model} can be equivalently re-written as
\begin{linenomath}
\begin{equation} \label{eq:model_with_facets}
 \mathbf{y}_{c,l} = \bs{\Phi}_{c,l} \overline{\mathbf{x}}_{c,l} + \mathbf{n}_{c,l},
\end{equation}
\end{linenomath}
where $\mathbf{y}_{c,l} \in \mathbb{C}^{\ndata_{c,l}}$ denotes the vector of $\ndata_{c,l}$ visibilities associated with the channel $l \in \{1, \ldots, \nband_c\}$, affected by a complex white Gaussian noise $\mathbf{n}_{c,l}$ with covariance matrix $\sigma^2_{c,l} \mathbf{I}_{\ndata_{c,l} \times \ndata_{c,l}}$. Introducing data blocks $b \in \{1, \ldots, \nblock\}$ as in~\eqref{eq:model_with_blocking}, the data model finally reads
\begin{linenomath}
\begin{equation} \label{eq:model_with_blocking_and_facets}
 \mathbf{y}_{c,l,b} = \bs{\Phi}_{c,l,b} \overline{\mathbf{x}}_{c,l} + \mathbf{n}_{c,l,b},
\end{equation}
\end{linenomath}
where $\mathbf{n}_{c,l} = (\mathbf{n}_{c,l,b})_{1 \leq b \leq \nblock}$, $\mathbf{y}_{c,l,b} \in \mathbb{C}^{\ndata_{c,l,b}}$ denotes the vector of $\ndata_{c,l,b}$ visibilities associated with the channel $l \in \{1, \ldots, \nband_c\}$ and data-block $b$, and $\mathbf{y}_{c,l} = (\mathbf{y}_{c,l,b})_{1 \leq b \leq \nblock}$. The operator $\bs{\Phi}_{c,l,b}$ and $\varepsilon_{c,l,b}$ are the associated measurement operator and $\ell_2$ ball radius, respectively, with $\bs{\Phi}_{c,l} = (\bs{\Phi}_{c,l,b})_{1 \leq b \leq \ncube}$. In this context, the initial minimization problem~\eqref{eq:problem} can be naturally substituted for
\begin{linenomath}
\begin{multline} \label{eq:problem-split-subcube}
	\minimize{\mathbf{X} \in \mathbb{R}_+^{\nbpix \times \nband}} 
	\sum_{c=1}^\ncube \Big( 
	\sum_{l=1}^{\nband_c} \sum_{b=1}^\nblock \iota_{\mathcal{B}(\mathbf{y}_{c,l,b}, \varepsilon_{c,l,b})} \bigl( \bs{\Phi}_{c,l,b}\mathbf{x}_{c,l} \bigr) \\
	+ r_c (\mathbf{X}_c)
	\Big),
\end{multline}
\end{linenomath}
where, for every $c \in \{1, \ldots, \ncube\}$, $\mathbf{X}_c = (\mathbf{x}_{c,l})_{1 \le l \le \nband_c} \in \mathbb{R}^{\nbpix \times \nband_c}$ is the $c$-th sub-cube of the full image cube $\mathbf{X}$, with $\mathbf{x}_{c,l} \in \mathbb{R}^\nbpix$ the $l$-th image of the sub-cube $\mathbf{X}_c$, and $r_c \colon \mathbb{R}^{\nbpix \times \nband_c} \to ]-\infty, +\infty]$ is a regularization term only acting on the $c$-th sub-cube. Note that SARA  and  HyperSARA correspond to the cases $(\ncube, \overline{\mu})=(\nband, 0)$ and $\ncube=1$, respectively.

%--------------------------------------------
\subsubsection{Spatial faceting} \label{sec:spatial_faceting}

Faceting can also be performed in the spatial domain by decomposing the regularization term for each spectral sub-cube into a sum of terms acting only locally in the spatial domain (see Figure~\ref{fig:facets} (c)). In this context, the resulting facets will need to overlap in order to mitigate edge effects, so that the overall objective function in~\eqref{eq:problem-split-subcube} will take the form of the sum of inter-dependent facet-specific objectives. Indeed, the splitting functionality of PDFB can be exploited to enable parallel processing of the facet-specific regularization terms and ensure additional degrees of parallelization compared to HyperSARA (see Section \ref{sec:algorithm}).
  
On the one hand, we propose to split the average sparsity dictionary $\bs{\Psi}^\dagger$ into $\nfacet$ smaller wavelet decomposition, leveraging the wavelet splitting technique introduced in~\citep[Chapter 4]{Prusa2012}. Indeed, \cite{Prusa2012} proposed an exact implementation of the discrete wavelet transform distributed over multiple facets. In this context, the Daubechies wavelet bases are decomposed into a collection of facet-based operators $\bs{\Psi}^\dag_q \in \mathbb{R}^{I_q \times \nbpix_q}$ acting only on the $q$-th facet of size $\nbpix_q$, with $I_q$ the associated number of wavelet coefficients and $I = I_1 + \ldots + I_\nfacet$. The overlap needed to ensure an exact faceted implementation of the SARA dictionary is composed of a number of pixels between $15(2^d - 2)$ and $15(2^d - 1)$ in each spatial direction~\citep[Section 4.1.4]{Prusa2012}, with $d$ the number of decomposition levels. In practice, the overlap ensures that each facet contains all the information needed to locally compute the convolutions underlying the discrete wavelet transforms.

On the other hand, we consider a faceted low-rankness prior enforced by the sum of logs priors acting on essentially the same overlapping facets as those introduced for the wavelet decomposition. This choice provides a more tractable alternative to the global low-rankness prior involved in HyperSARA. Unlike the wavelet decomposition, there is no equivalent faceted implementation of the SVD. To mitigate any potential reconstruction artifacts resulting from the faceting of the 3D image cube, we propose to consider a larger facet size $\widetilde{\nbpix}_q$ ($q\in \{1, \ldots, \nfacet\}$) for these terms, and introduce diagonal matrices $ \mathbf{D}_q \in ]0,+\infty[^{\widetilde{\nbpix}_q \times \widetilde{\nbpix}_q}$ to ensure a smooth transition from the borders of one facet to its neighbours. A natural choice consists in down-weighting the contribution of pixels involved in multiple facets. A tapering window, ensuring amplitude decay for pixels involved in multiple contiguous facets while imposing the sum of all the weights associated with each pixel to be equal to unity, is considered. In this work, we consider weights in the form of a 2D triangular apodization window as considered by~\citet{Murya2017} (see Figure~\ref{fig:facets} (c)), defined as the outer product of two 1D piece-wise linear apodization windows. The amount of overlap for this regularization term is taken as an adjustable parameter of the Faceted HyperSARA approach. Its influence is investigated in Section~\ref{sec:exp_simu}.

The spatial faceting procedure enables the splitting of the original log-sum priors~\eqref{eq:hypersara0} of HyperSARA into a sum of inter-dependent, facet-specific log-sum priors underpinning the Faceted HyperSARA prior, which reads
\begin{linenomath}
\begin{multline} \label{eq:log-sum:FHsara}
 r_c(\mathbf{X}_c)
 = \sum_{q=1}^{\nfacet} 
 \Bigg( \overline{\mu}_{c,q} \sum_{j=1}^{J_{c,q}} \overline{\upsilon}_{c, q} \log \bigg( \frac{| s_j( \mathbf{D}_q \widetilde{\mathbf{S}}_q \mathbf{X}_c ) |}{\overline{\upsilon}_{c,q}} + 1 \bigg) \\
 + \mu_c \sum_{i=1}^{I_q} \upsilon_c \log \bigg( \frac{\| [ \bs{\Psi}_q^\dag \mathbf{S}_q \mathbf{X}_c ]_i \|_2}{\upsilon_c} + 1 \bigg) \Bigg),
\end{multline}
\end{linenomath}
where, for every $q \in \{1, \ldots, \nfacet\}$, 
$(\overline{\mu}_{c, q},  \overline{\upsilon}_{c,q}, \mu_c,\upsilon_c) \in ]0,+\infty[^4$ are regularization parameters, $J_{c,q} = \min\{\widetilde{\nbpix}_q, \nband_c\}$, and $\nwcoefs_q$ is the number of wavelet coefficients generated by the operator $\bs{\Psi}^\dagger_q$. Note that the parameters $\mu_c$, $\upsilon_c$ do not depend on the facet index $q$, since the faceted joint average sparsity prior is an exact reformulation of the sparsity prior leveraged in HyperSARA (see~\eqref{eq:hypersara0}). The linear operators $\widetilde{\mathbf{S}}_q \in \mathbb{R}^{\widetilde{\nbpix}_q \times \nbpix}$ and $\mathbf{S}_q \in \mathbb{R}^{\nbpix_q \times \nbpix}$ extract spatially overlapping spatio-spectral facets from the image cube $\mathbf{X}_c$ for the respective low-rankness and average sparsity priors, of respective sizes $\widetilde{\nbpix}_q$ and $\nbpix_q$. These two operators only differ in the amount of overlapping pixels considered, which is defined as an adjustable parameter for $\widetilde{\mathbf{S}}_q$, and prescribed by~\citet{Prusa2012} for $\mathbf{S}_q$. Each facet relies on a spatial decomposition of the image into non-overlapping tiles (see Figure~\ref{fig:facets}~(b), delineated by dashed red lines), overlapping with its top and left spatial neighbour. In the following, the overlapping regions will be referred to as the borders of a facet, in contrast with its underlying tile (see Figure~\ref{fig:facets}). An edge facet, \emph{i.e.} which does not have a neighbour in one of the two spatial dimensions, has the same dimension as the underlying tile in the direction where it does not admit a neighbour (\emph{e.g.} facets in the left-hand side corners have the same dimension as the underlying tile). Note that SARA and HyperSARA correspond to the cases $(\nfacet, \ncube, (\overline{\mu}_{c, q})_{c, q}) = (1, \nband, \mathbf{0}_{\ncube \times \nfacet})$ and $(\nfacet, \ncube) = (1, 1)$, respectively. 

The reweighting approach used to minimize the objective \eqref{eq:problem-split-subcube} with the log-sum priors \eqref{eq:log-sum:FHsara} via convex relaxations powered by PDFB is described in Section~\ref{sec:algorithm}. Crucially, the splitting functionality of PDFB will be exploited to enable parallel processing of these facet-specific priors, more specifically their convex relaxations. 

\paragraph*{Heuristic for automatic regularization parameter setting with Faceted HyperSARA.}
We generalize the HyperSARA heuristic, building on the same observation that $\mu_c$ and $\overline{\mu}_{c,q}$ act as soft-thresholding parameters in the wavelet and singular value domains respectively (explicit formulas to appear in Section \ref{sec:algorithm} Algorithm~\ref{alg:pdfb} lines \ref{alg:step:dual_lr} and \ref{alg:step:dual_sparsity}, and Table \ref{tab:proximity}), to be set to corresponding noise level values, similarly to $\upsilon_c$ and $\overline{\upsilon}_{c,q}$. In the wavelet domain, the faceted wavelet transform being exactly equal to the non-faceted one, spatial faceting has no impact on the noise level. The heuristic \eqref{eq:heuristic_hypersara_sparsity} thus trivially generalizes for each sub-cube $c$ (associated with the spectral faceting) to
\begin{linenomath}
\begin{equation} \label{eq:heuristic_fhs_sparsity}
    \mu_c = \upsilon_c = \overline{\sigma}_c[m^{(\chi_{\nband_c})} + s^{(\chi_{\nband_c})}],
\end{equation}
\end{linenomath}
with 
$\overline{\sigma}^2_c=1/\nband_c
    \sum_{l=1}^{\nband_c} \overline{\sigma}^2_{c,l}$, and $\overline{\sigma}_{c,l}=\sigma_{c,l}/\norm{\bs{\Phi}_{c,l}}_\text{S}$.
For the singular value domain, the heuristic \eqref{eq:heuristic_hypersara_low_rank} generalizes for each spatial facet $q$ of sub-cube $c$ to:
\begin{linenomath}
\begin{equation} \label{eq:heuristic_fhs_low_rank}
    \overline{\mu}_{c,q} = \overline{\upsilon}_{c,q} = \overline{\sigma}_c\sqrt{\frac{\widetilde{\nbpix}_q \nband_c}{J_{c,q}}}.
\end{equation}
\end{linenomath}

%% ==============================================================
%% ALGORITHM
%% ==============================================================
\section{Faceted HyperSARA algorithm} \label{sec:algorithm}

The parallel algorithmic structure of Faceted HyperSARA is described in this section, leveraging PDFB within a reweighting approach to handle the log-sum priors. Implementation details are also discussed.

\begin{algorithm}[t]
\small
	\KwData{$(\mathbf{y}_{c,l,b} )_{c,l,b}$, $c \in \{1, \dotsc, C\}$, $l \in \{1, \dotsc, \nband_c \}$, $b \in \{1, \dotsc, \nblock \}$}
	\KwIn{$\mathbf{X}^{(0)} = (\mathbf{X}_c^{(0)})_c$, $\mathbf{P}^{(0)} = (\mathbf{P}_c^{(0)})_c$, $\mathbf{W}^{(0)} = (\mathbf{W}_c^{(0)})_c$, $\mathbf{v}^{(0)} = (\mathbf{v}_c^{(0)})_c$}
	
	\Parameter{$\nreweightmax > 0$, $0 < \relvarrw < 1$}
  	
  	$t \leftarrow 0$, $\xi \leftarrow +\infty$\;
    
    \tcp{Solve spectral sub-problems in parallel}
	\For{$c = 1$ \KwTo $C$ \label{algo:rw:forC}}{
  	\While{$(t < \nreweightmax)$ and $(\xi > \relvarrw)$}{
	
	    \For{$q = 1$ \KwTo $Q$}{
    		\tcp{Update weights (low-rankness prior)}
    		$\overline{\bs{\theta}}_{c,q}^{(t)} =
    		    \overline{\bs{\omega}}_{c,q}(\mathbf{X}_c^{(t)})$; \quad \tcp{using \eqref{eq:facet_hypersara_rank}} \label{alg:step:weights_nuclear}
		    
    		\tcp{Update weights (joint-sparsity prior)}
    		$\bs{\theta}_{c,q}^{(t)} =
    		    \bs{\omega}_{c,q}(\mathbf{X}_c^{(t)})$; \quad \tcp{using \eqref{eq:facet_hypersara_sparsity}}
    		    \label{alg:step:weights_l21}
    	}
  	
  		\tcp{Run PDFB (Algorithm~\ref{alg:pdfb})}
  	    $(\mathbf{X}_c^{(t+1)}, \mathbf{P}_c^{(t+1)}, \mathbf{W}_c^{(t+1)}, \mathbf{v}_c^{(t+1)}) = $\nonl\\
  	    $ \qquad \textbf{Algorithm2}\bigl(\mathbf{X}_c^{(t)}, \mathbf{P}_c^{(t)}, \mathbf{W}_c^{(t)}, \mathbf{v}_c^{(t)}, \bs{\theta}_c^{(t)}, \overline{\bs{\theta}}_c^{(t)}  \bigr) $\; \label{alg:step:pdfb}
    }
    $\xi = \norm{\mathbf{X}_c^{(t+1)}-\mathbf{X}_c^{(t)}}_\text{F}/\norm{\mathbf{X}_c^{(t)}}_\text{F}$\;
	$t \leftarrow t+1$\;
	}	
	\KwResult{$\mathbf{X}^{(t)}$, $\mathbf{P}^{(t)}$, $\mathbf{W}^{(t)}$, $\mathbf{v}^{(t)}$}
	\caption{Outer reweighting algorithm.}
	\label{alg:reweighting}
\end{algorithm}

%--------------------------------------------
\subsection{Outer reweighting algorithm}
\label{ssec:algorithm-description}

To efficiently address the log-sum prior underpinning the Faceted HyperSARA prior, we resort to a majorize-minimize algorithm similar to the one proposed by~\citep{Candes2009}, leading to the reweighting approach described in Algorithm~\ref{alg:reweighting}.

At each iteration $t \in \mathbb{N}$ of Algorithm~\ref{alg:reweighting}, problem~\eqref{eq:problem-split-subcube} is majorized at the local estimate $\mathbf{X}^{(t)}$ by a convex approximation, and then minimized using a PDFB algorithm described in Algorithm~\ref{alg:pdfb} (see Algorithm~\ref{alg:reweighting} line~\ref{alg:step:pdfb}). 
For each sub-cube $c \in \{1, \ldots, \ncube\}$, the convex approximated minimization problem is of the form

\begin{linenomath}
\begin{multline} \label{eq:problem-split-subcube-approx}
	\minimize{\mathbf{X}_c \in \mathbb{R}_+^{\nbpix \times \nband}} 
	\sum_{l=1}^{\nband_c} \sum_{b=1}^\nblock \iota_{\mathcal{B}(\mathbf{y}_{c,l,b}, \varepsilon_{c,l,b})} \bigl( \bs{\Phi}_{c,l,b}\mathbf{x}_{c,l} \bigr) \\
	+ \widetilde{r}_c (\mathbf{X}_c, \mathbf{X}^{(t)}_c),
\end{multline}
\end{linenomath}
where $\widetilde{r}_c( \cdot, \mathbf{X}_c^{(t)})$ is a convex local majorant function of $r_c$ at $\mathbf{X}_c^{(t)}$, corresponding to the weighted hybrid norm prior
\begin{linenomath}
\begin{multline} \label{eq:facet_hypersara}
	\widetilde{r}_c(\mathbf{X}_c, \mathbf{X}^{(t)}_c) = \sum_{q=1}^\nfacet 
	\Big( \overline{\mu}_{c, q} \norm{\mathbf{D}_q \widetilde{\mathbf{S}}_q \mathbf{X}_c}_{\ast, \overline{\bs{\omega}}_{c,q}(\mathbf{X}_c^{(t)})} \\
	+ \mu_c \norm{\bs{\Psi}_q^\dag \mathbf{S}_q \mathbf{X}_c}_{2, 1, \bs{\omega}_{c,q}(\mathbf{X}_c^{(t)})} \Big).%
\end{multline}
\end{linenomath}
For every $q \in \{ 1, \dotsc, \nfacet \}$, the weights $\overline{\bs{\omega}}_{c,q}(\mathbf{X}_c^{(t)})=\big( \overline{\omega}_{c,q,j}(\mathbf{X}_c^{(t)}) \big)_{1 \le j \le J_{c,q}} $ and $\bs{\omega}_{c,q}(\mathbf{X}_c^{(t)}) = \big( \omega_{c,q,i}(\mathbf{X}_c^{(t)}) \big)_{1 \le i \le I_q} $ are given by
\begin{linenomath}
\begin{equation}
  \overline{\omega}_{c,q,j}(\mathbf{X}_c^{(t)})
 = \overline{\upsilon}_{c, q} \Big( \big| s_j (\mathbf{D}_q \widetilde{\mathbf{S}}_q \mathbf{X}^{(t)}_c) \big| + \overline{\upsilon}_{c, q} \Big)^{-1}, \; \overline{\upsilon}_{c, q} > 0,  \label{eq:facet_hypersara_rank} 
\end{equation}
\end{linenomath}
\begin{linenomath}
\begin{equation}
    \omega_{c,q,i}(\mathbf{X}_c^{(t)})
 = \upsilon_c \Big( \| [\bs{\Psi}_q^\dag \mathbf{S}_q \mathbf{X}^{(t)}_c]_i \|_2 + \upsilon_c \Big)^{-1},\; \upsilon_c > 0, \label{eq:facet_hypersara_sparsity}
\end{equation}
\end{linenomath}
and the weighted norms $\norm{\cdot}_{\ast, \overline{\bs{\omega}}_{c,q}(\mathbf{X}_c^{(t)})}$ and $\norm{\cdot}_{2, 1, \bs{\omega}_{c,q}(\mathbf{X}_c^{(t)})}$ are respectively defined by
\begin{linenomath}
\begin{multline}
    (\forall \mathbf{Z} \in \mathbb{R}^{\widetilde{\nbpix}_q \times \nband_c}), \\ \norm{\mathbf{Z}}_{\ast, \overline{\bs{\omega}}_{c,q}(\mathbf{X}_c^{(t)})} = \sum_{j=1}^{J_{c,q}} \overline{\omega}_{c,q,j}(\mathbf{X}_c^{(t)}) s_j(\mathbf{Z}), \label{eq:weighted_nuclear_norm}
\end{multline}
\end{linenomath}
\begin{linenomath}
\begin{multline}
(\forall \mathbf{Z} \in \mathbb{R}^{\nwcoefs_q \times \nband_c} ), \\ \norm{\mathbf{Z}}_{2, 1, \bs{\omega}_{c,q}(\mathbf{X}_c^{(t)})} = \sum_{i=1}^{\nwcoefs_q} \omega_{c,q, i}(\mathbf{X}_c^{(t)}) \| [\mathbf{Z}]_i \|_2. \label{eq:weighted_l21_norm} 
\end{multline}
\end{linenomath}

The sequence $(\mathbf{X}^{(t)})_t = \Big( \big(\mathbf{X}^{(t)}_c \big)_{1\leq c \leq \ncube} \Big)_t$ is assured to converge to a critical point of the non-convex problem~\eqref{eq:problem-split-subcube}, given that the solution provided by Algorithm~\ref{alg:pdfb} is accurate enough (see \emph{e.g.} \citet{Ochs2015}). A complete description of the PDFB-powered algorithm used to solve the sub-problems~\eqref{eq:problem-split-subcube-approx} is provided in the next section.

From~\eqref{eq:facet_hypersara_rank} and~\eqref{eq:facet_hypersara_sparsity}, note that the weights are normalized in $]0,1]$. Large weights are obtained in locations where the estimate from the previous iteration is small compared to the floor level, inducing a significant penalization in~\eqref{eq:weighted_nuclear_norm} and~\eqref{eq:weighted_l21_norm}. On the contrary, large signal values with respect to the floor level will induce weights close to zero, barely penalizing signal values in the objective. A closer look into the PDFB-powered algorithm to solve the sub-problems~\eqref{eq:problem-split-subcube-approx} described in Section \ref{ssec:algorithm:overview} (see Algorithm~\ref{alg:pdfb} lines \ref{alg:step:dual_lr} and \ref{alg:step:dual_sparsity}) will reveal that the soft-thresholding parameters are in fact not $\mu_c$ and $\overline{\mu}_{c,q}$ themselves but their product with the weights. Large weights (close or equal to 1) will preserve the soft-threshold value at the floor level (set to the noise level according to the heuristic), while small weights will simply drive it close to zero, making the soft-thresholding inactive, thus preserving the signal value.

\begin{table*}
    \centering
    \caption{Proximal operators involved in Algorithm~\ref{alg:pdfb}.}
    \begin{tabular}{@{}ll@{}}\toprule
        \textbf{Proximal operator for} $\alpha>0$ & \textbf{Description} \\ \midrule
        $\prox_{\alpha \norm{\cdot}_1} (\mathbf{z}) = \sgn(\mathbf{z}) \max \bigl\{\lvert \mathbf{z} \rvert - \alpha, \mathbf{0} \bigr\}$ & $\mathbf{z} \in \mathbb{R}^\nbpix$, soft-thresholding operator \\
        $\prox_{\alpha \norm{\cdot}_{*}} (\mathbf{Z}) = \mathbf{U} \Diag \bigr(\prox_{\alpha \norm{\cdot}_1} (\bs{\sigma}) \bigl) \mathbf{V}^\dag$ & $\mathbf{Z} = \mathbf{U}\bs{\Sigma}\mathbf{V}^\dag \in \mathbb{R}^{\nbpix \times \nband}$, $\bs{\Sigma} = \Diag(\bs{\sigma})$\\
        & singular value decomposition of $\mathbf{Z}$\\
        $\prox_{\alpha \norm{\cdot}_{2, 1}} (\mathbf{Z}) = \Bigl( \max \bigl\{ \norm{\mathbf{z}_n^\dagger}_2 - \alpha, 0 \bigr\} \frac{\mathbf{z}_n^\dagger}{\norm{\mathbf{z}_n^\dagger}_2} \Bigr)_{1 \leq n \leq \nbpix}$ & $\mathbf{Z} = [\mathbf{z}_1, \dotsc, \mathbf{z}_\nbpix]^\dag \in \mathbb{R}^{\nbpix \times \nband}$\\[0.25cm]
        $\prox_{\iota_{\mathcal{B}(\mathbf{y}_{c,l,b}, \varepsilon_{c,l,b})}} (\mathbf{z}) 
        = \dfrac{\mathbf{z} - \mathbf{y}_{c,l,b}}{\max \big\{ \| \mathbf{z} - \mathbf{y}_{c,l,b} \|_2/\varepsilon_{c,l,b}, 1 \big\} } + \mathbf{y}_{c,l,b}$
        & $\mathbf{z} \in \mathbb{C}^{M_{c,l,b}}$\\[0.25cm]
        $\prox_{\iota_{\mathbb{R}^{\nbpix \times \nband}_+}} (\mathbf{Z}) = \max \bigl\{\mathbf{0}, \Re(\mathbf{Z}) \big\}$ & $\mathbf{Z} \in \mathbb{C}^{\nbpix \times \nband}$, $\Re(\mathbf{Z})$ denotes the real part of $\mathbf{Z}$\\ \bottomrule
    \end{tabular}
    \label{tab:proximity}
\end{table*}

\begin{algorithm}[tp]
\small
	\KwData{$(\mathbf{y}_{c,l,b} )_{l,b}$, $l \in \{1, \dotsc, \nband_c \}$, $b \in \{1, \dotsc, \nblock \}$\medskip}
	\KwIn{$\mathbf{X}_c^{(0)}$, $\mathbf{P}_c^{(0)} = \bigl( \mathbf{P}_{c,q}^{(0)} \bigr)_q$, $\mathbf{W}_c^{(0)} = \bigl( \mathbf{W}_{c,q}^{(0)} \bigr)_{q}$, $\mathbf{v}_c^{(0)} = \bigl( \mathbf{v}_{c,l,b}^{(0)} \bigr)_{l,b}$, $\bs{\theta}_c = \big( \bs{\theta}_{c,q} \big)_{1 \le q \le \nfacet}$, $\overline{\bs{\theta}}_c = \big( \overline{\bs{\theta}}_{c,q} \big)_{1 \le q \le \nfacet}$\medskip}
	
	\Parameter{$(\mathbf{D}_{q})_q$, $( \mathbf{U}_{c,l,b})_{l,b}$, $\bs{\varepsilon} = (\varepsilon_{c,l,b})_{l,b}$, $\mu_c$, $(\overline{\mu}_{c, q})_q$, $\tau$, $\zeta$, $(\eta_{c,l})_{1 \leq l \leq \nband}$, $\kappa$, $0 < \npdfbmin < \npdfbmax$, $0 < \relvarpdfb < 1$ \medskip}
	
  	$p \leftarrow 0$; $\xi = +\infty$;
    $\check{\mathbf{X}}_c^{(0)} = \mathbf{X}_c^{(0)}$, $\hat{\mathbf{X}}_c^{(0)} = (\hat{\mathbf{x}}_{c,l}^{(0)})_{1 \le l \le L_c} = \mathbf{FZ}\mathbf{X}_c^{(0)}$\;
    $\mathbf{r}_c^{(0)} = (r_{c,l,b}^{(0)})_{l,b} \in \mathbb{R}^{\nband_c \nblock}$, with $r_{c,l,b}^{(0)} = \norm{\mathbf{y}_{c,l,b} - \bs{\Phi}_{c,l,b} \mathbf{x}_{c,l}^{(0)}}_2$\;
  	
  	\While{$(p < \npdfbmin)$ or $\big[ (p < \npdfbmax)$ and $(\xi > \relvarpdfb $ or $\|\mathbf{r}_c^{(p)}\|_2 > 1.01 \|\bs{\varepsilon}\|_2)\big]$}{

		\tcp{Update low-rankness and sparsity variables}
		
  		split $(\widetilde{\mathbf{X}}_{c,q}^{(p)})_{1 \le q \le \nfacet} = (\widetilde{\mathbf{S}}_q\check{\mathbf{X}}_c^{(p)})_{1 \le q \le \nfacet}$\; \label{alg:eq:broadcast_facet}
		
  		split $(\check{\mathbf{X}}_{c,q}^{(p)})_{1\le q \le \nfacet} = (\mathbf{S}_q \check{\mathbf{X}}_c^{(p)})_{1\le q \le \nfacet}$\;
  		
		\tcp{[Parallel on facet cores]}
		\For{q = 1 \KwTo \nfacet \label{alg:step:start_facet-nuclear}}{
		$\mathbf{P}_{c,q}^{(p+1)} = \bigl( \mathbf{I}_{\widetilde{N}_q \times \widetilde{N}_q} - \prox_{\zeta^{-1} \overline{\mu}_{c, q} \lVert \cdot \rVert_{\ast,\overline{\bs{\theta}}_{c,q} }} \bigr) \bigl( \mathbf{P}_{c,q}^{(p)} + \mathbf{D}_{q} \widetilde{\mathbf{X}}_{c,q}^{(p)} \bigr)$\; \label{alg:step:dual_lr}
		$\widetilde{\mathbf{P}}_{c,q}^{(p+1)} = \mathbf{D}_{q}^\dag \mathbf{P}_c^{(p+1)}$\; \label{alg:step:end_facet-nuclear}

		$\mathbf{W}_{c,q}^{(p+1)} = \bigl( \mathbf{I}_{\nwcoefs{}_q \times \nwcoefs{}_q} - \prox_{\kappa^{-1}\mu_c \lVert \cdot \rVert_{2, 1,\bs{\theta}_{c,q}}} \bigr) \bigl( \mathbf{W}_{c,q}^{(p)} + \bs{\Psi}^\dag_q \check{\mathbf{X}}_{c,q}^{(p)} \bigr)$\; \label{alg:step:dual_sparsity}
		$\widetilde{\mathbf{W}}_{c,q}^{(p+1)} = \bs{\Psi}_q \mathbf{W}_{c,q}^{(p+1)}$\; \label{alg:step:end_facet-l21} 
		}	
		
	    \tcp{Update data fidelity variables [data cores]}
  		\For{$l$ = 1 \KwTo $\nband_c$}{
  		    $\hat{\mathbf{x}}^{(p+1)}_{c,l} = \mathbf{FZ} {\mathbf{x}}_{c,l}^{(p)}$\label{alg:eq:fourier}
  		    
  		    split $(\hat{\mathbf{x}}^{(p+1)}_{c,l,b})_{1 \le b \le \nblock} = (\mathbf{M}_{c,l,b} \hat{\mathbf{x}}^{(p+1)}_{c,l})_{1 \le b \le \nblock}$\;
  		    \label{alg:eq:broadcast_fourier}

		\For{ $b$ = 1 \KwTo $\nblock$ \label{alg:step:start_data}}{
    		$\mathbf{v}_{c,l,b}^{(p+1)} = \mathbf{U}_{c,l,b} \bigl( \mathbf{I}_{M_{c,l,b}} - \prox^{\mathbf{U}_{c,l,b}}_{\iota_{\mathcal{B}(\mathbf{y}_{c,l,b}, \varepsilon_{c,l,b})}} \bigr) \bigl($\nonl \\  
    		\hfill $ \mathbf{U}_{c,l,b}^{-1}\mathbf{v}_{c,l,b}^{(p)} + \, \mathbf{G}_{c,l,b} (2\hat{\mathbf{x}}_{c,l,b}^{(p+1)}-\hat{\mathbf{x}}_{c,l,b}^{(p)}) \bigr)$\; \label{alg:step:dual_data1}
    		
    		$\widetilde{\mathbf{v}}_{c,l,b}^{(p+1)} = \mathbf{G}_{c,l,b}^\dag \mathbf{v}_{c,l,b}^{(p+1)}$\; \label{alg:step:dual_data2}
            $r_{c,l,b}^{(p+1)} = \norm{\mathbf{y}_{c,l,b} - \mathbf{G}_{c,l,b} \hat{\mathbf{x}}_{c,l,b}^{(p+1)}}_2$\;\label{alg:step:residual}
		}
  		}

		\tcp{Inter node communications}
		\For{l = 1 \KwTo $\nband_c$}{	 
		$\displaystyle \mathbf{a}_{c,l}^{(p)} = \sum_{q=1}^\nfacet \Bigl( \zeta \widetilde{\mathbf{S}}_q^\dag\widetilde{\mathbf{p}}_{c,q,l}^{(p+1)}
			 + \kappa \mathbf{S}_q^\dag \widetilde{\mathbf{w}}_{c,q,l}^{(p+1)} \Bigr) $ \nonl\\
			 $\displaystyle \hfill + \, \eta_{c,l} \mathbf{Z}^\dag \mathbf{F}^\dag \sum_{b} \mathbf{M}_{c,l,b}^\dag \widetilde{\mathbf{v}}_{c,l,b}^{(p+1)} $\; \label{alg:step:aggregate}
	    }
			 
	    \tcp{Update image tiles [on facet cores, in parallel]}
  		$\mathbf{X}_c^{(p+1)} = \prox_{\iota_{\mathbb{R}^{\nbpix \times \nband_c}_+}} \bigl( \mathbf{X}_c^{(p)} - \tau \mathbf{A}_c^{(p)}  \bigr)$\label{alg:step:primal}\tcp*{$\mathbf{A}_c^{(p)} = \big(\mathbf{a}_{c,l}^{(p)} \big)_{1\leq l \leq \nband}$}
  		$\check{\mathbf{X}}_c^{(p+1)} = 2\mathbf{X}_c^{(p+1)} - \mathbf{X}_c^{(p)}$\tcp*{communicate facet borders}

		$\xi = \norm{\mathbf{X}_c^{(p+1)}-\mathbf{X}_c^{(p)}}_\text{F}/\norm{\mathbf{X}_c^{(p)}}_\text{F}$\;
		$p \leftarrow p+1$\;
    }
	\KwResult{
	$\mathbf{X}_c^{(p)}$, $\mathbf{P}_c^{(p)}$, $\mathbf{W}_c^{(p)}$, $\mathbf{v}_c^{(p)}$}
	\caption{
	\label{alg:pdfb}
	Inner convex optimisation algorithm (PDFB) to solve each spectral sub-problem \eqref{eq:problem-split-subcube-approx}.
	}
\end{algorithm}

%--------------------------------------------
\subsection{Inner convex optimization algorithm} \label{ssec:algorithm:overview}

A primal-dual algorithmic structure as PDFB jointly solves the problem~\eqref{eq:problem-split-subcube-approx}, referred to as the primal problem, and its dual formulation in the sense of the Fenchel-Rockafellar duality theory~\citep{Bauschke2017book}. The dual problem of~\eqref{eq:problem-split-subcube-approx} is expressed as
\begin{linenomath}
\begin{align} \label{eq:problem-split-subcube-approx-dual}
\begin{split}
	&\minimize{(\mathbf{P}_{c,q})_q, (\mathbf{W}_{c,q})_q, (\mathbf{v}_{c,l,b})_{l, b}}
	\sum_{l=1}^{\nband_c} \sum_{b=1}^\nblock \iota_{\mathcal{B}(\mathbf{y}_{c,l,b}, \varepsilon_{c,l,b})}^* ( \mathbf{v}_{c,l,b}) \\
	&+ \iota_{\mathbb{R}_+^{\nbpix \times \nband}}^* \Big( 
	    - \sum_{q=1}^\nfacet \big( \widetilde{\mathbf{S}}_q^\dagger \mathbf{D}_{q}^\dagger \mathbf{P}_{c,q} + \mathbf{S}_q^\dagger \mathbf{W}_{c,q} \big)
	    - \sum_{l=1}^{\nband_c} \sum_{b=1}^\nblock \bs{\Phi}_{c,l,b}^\dagger \mathbf{v}_{c,l,b} \Big) \\
	&+ \sum_{q=1}^\nfacet f_{c,q,t}^* (\mathbf{P}_{c,q}) + g_{c,q,t}^* (\mathbf{W}_{c,q}),
\end{split}
\end{align}
\end{linenomath}
where $f_{c,q,t} = \overline{\mu}_{c, q} \norm{\cdot}_{\ast, \overline{\bs{\omega}}_{c,q}(\mathbf{X}_c^{(t)})}$, $g_{c,q,t} = \mu_c \norm{\cdot}_{2, 1, \bs{\omega}_{c,q}(\mathbf{X}_c^{(t)})}$, and the superscript $*$ denotes the convex conjugate of a function~\citep[Chapter 13]{Bauschke2017book}. The variables $\mathbf{P}_{c,q} \in \mathbb{R}^{\widetilde{\nbpix}_q \times \nband_c}$, $\mathbf{W}_{c,q} \in \mathbb{R}^{\nbpix_q \times \nband_c}$ and $\mathbf{v}_{c,l,b} \in \mathbb{C}^{\ndata_{c,l,b}}$ are referred to as the \emph{dual variables}. The splitting functionality will enable all block-specific data-fidelity terms and facet-specific regularization terms to be updated in parallel via their proximal operator. In this work, we resort to a preconditioned variant of PDFB, which uses proximal operators with respect to non-Euclidean metrics in order to reduce the number of iterations necessary for the algorithm to converge. Let $\mathbf{U} \in \mathbb{R}^{n \times n}$ be a symmetric, positive definite matrix. The proximal operator of a proper, convex, lower semi-continuous function $f \colon \mathbb{R}^n \rightarrow ]-\infty,+\infty]$ at $\mathbf{z} \in \mathbb{R}^n$ with respect to the metric induced by $\mathbf{U}$ is defined by~\citep{Moreau1965, Hiriart1993}
\begin{linenomath}
\begin{equation} \label{eq:prox_definition}
 \prox_f^\mathbf{U} (\mathbf{z}) = \argmind{\mathbf{x} \in \mathbb{R}^n} \Bigl\{ f(\mathbf{x}) + \frac{1}{2} (\mathbf{x} - \mathbf{z})^\dag \mathbf{U} (\mathbf{x} - \mathbf{z}) \Bigr\}.
\end{equation}
\end{linenomath}
The more compact notation $\prox_f$ is used when $\mathbf{U} = \mathbf{I}_{n \times n}$, where $\mathbf{I}_{n \times n}$ is the identity matrix in $\mathbb{R}^{n \times n}$. In addition, when the function $f$ corresponds to an indicator function of a closed, non-empty, convex set, its proximal operator reduces to a projection operator.

\begin{figure*}
	\centering
	\includegraphics[keepaspectratio,width=0.75\textwidth]{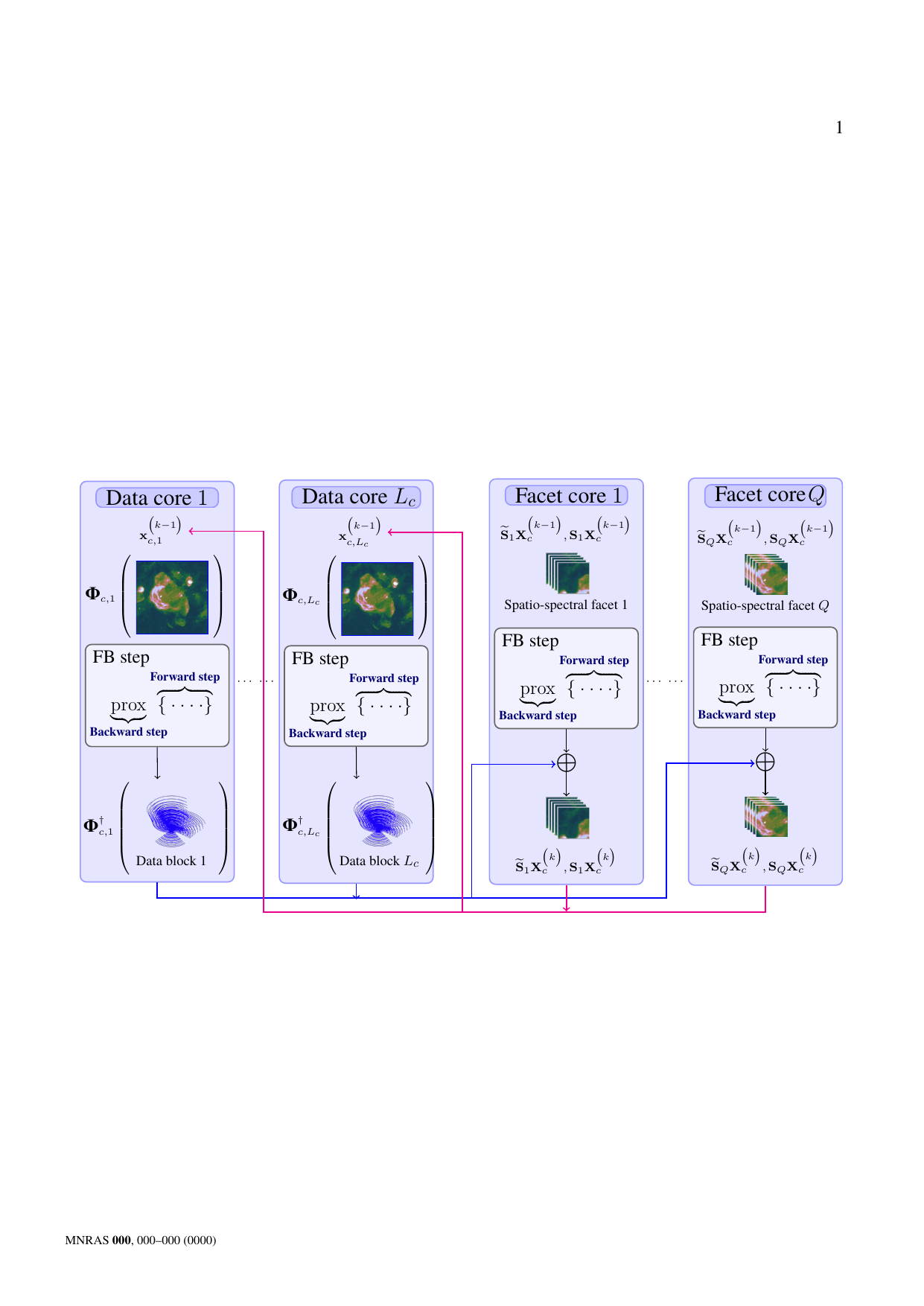}
    \caption{Illustration of the two groups of cores described in Section~\ref{sec:algorithm}, with the main steps involved in Algorithm~\ref{alg:pdfb} applied to each independent sub-problem $c \in \{1, \dotsc, \ncube\}$, using $\nfacet$ facets (along the spatial dimension) and $\nblock = 1$ data block per channel. Data cores handle variables of the size of data blocks (Algorithm~\ref{alg:pdfb} lines~\ref{alg:step:start_data}--\ref{alg:step:dual_data2}), whereas facet cores handle variables of the size of a spatio-spectral facet (Algorithm~\ref{alg:pdfb} lines~\ref{alg:step:start_facet-nuclear}--\ref{alg:step:end_facet-l21}). Communications between the two groups are represented by colored arrows. Communications between facet cores, induced by the overlap between the spatio-spectral facets, are illustrated in Figure~\ref{fig:facet_communications}.}
    \label{fig:algo}
\end{figure*}

A graphical illustration of the PDFB-powered algorithm to solve problem~\eqref{eq:problem-split-subcube-approx} is given in Figure~\ref{fig:algo}. A formal description is reported in Algorithm~\ref{alg:pdfb}. The expression of the different proximal operators involved in Algorithm~\ref{alg:pdfb} are given in Table~\ref{tab:proximity}.

First, the faceted low-rankness prior is handled in lines~\ref{alg:step:dual_lr}-\ref{alg:step:end_facet-nuclear} by computing in parallel the proximal operator of the per-facet weighted nuclear norms. Second, the average sparsity prior is addressed in lines~\ref{alg:step:dual_sparsity}-\ref{alg:step:end_facet-l21} by computing the proximal operator of the per-facet weighted $\ell_{2,1}$ norm in parallel. Third, the data-fidelity terms are handled in parallel in lines~\ref{alg:step:start_data}-\ref{alg:step:dual_data2} by computing, for every data block $(c,l,b)$, the projection onto the $\ell_2$ balls $\mathcal{B}(\mathbf{y}_{c,l,b}, \varepsilon_{c,l,b})$ with respect to the metric induced by the diagonal matrices $\mathbf{U}_{c,l,b}$, chosen using the preconditioning strategy proposed by~\citet{Onose2017, Abdulaziz2019}.
More precisely, $\mathbf{U}_{c,l,b}$ is a diagonal matrix, whose coefficients are the inverse of the sampling density in the vicinity of the acquired Fourier modes. 
The projection onto the $\ell_2$ ball for the metric induced by $\mathbf{U}_{c,l,b}$ does not admit an analytic expression, and thus needs to be approximated numerically through sub-iterations. In this work, we adopt a forward-backward algorithm~\citep{Combettes2011}  
which iteratively approximates this projection by computing Euclidean projections onto the $\ell_2$ ball $\mathcal{B}(\mathbf{y}_{c,l,b}, \varepsilon_{c,l,b})$ \citep[see][Algorithm 2]{Onose2017}.
Finally, the non-negativity constraint is handled in line~\ref{alg:step:primal} by computing the Euclidean projection onto the non-negative orthant.

For a given sub-cube $c \in \{1, \dotsc, \ncube \}$, Algorithm~\ref{alg:pdfb} is guaranteed to converge to a global solution to the sub-problem~\eqref{eq:problem-split-subcube-approx}, provided that the parameters $(\tau, \zeta, \eta, \kappa)$ satisfy the conditions described in \citep[Lemma 4.3]{Pesquet2014}.
In particular, these conditions are satisfied for the following choice of parameters $\tau = 0.33$, $\zeta = 1$, $\eta_{c,l} = 1/ \norm{\mathbf{U}_{c,l}^{1/2}\bs{\Phi}_{c,l}}^2_\text{S}$, $ \kappa = 1/\norm{\bs{\Psi}^\dag}^2_\text{S}$.

In summary, at each iteration $p$ of the reweighting scheme (Algorithm~\ref{alg:reweighting}), the PDFB algorithm (Algorithm~\ref{alg:pdfb}) yields, at convergence, a minimizer $\mathbf{X}_c^{(t)}$ of the convex problem~\eqref{eq:problem-split-subcube-approx}. The sequence $(\mathbf{X}^{(t)})_t = \Big( \big(\mathbf{X}^{(t)}_c \big)_{1\leq c \leq \ncube} \Big)_t$ is itself assured to converge to a critical point of the global non-convex problem~\eqref{eq:problem-split-subcube}.

Note that PDFB can accommodate randomization in the update of the variables, \emph{e.g.} by randomly selecting a subset of the data and facet dual variables to be updated at each iteration. This procedure can significantly alleviate the memory load per node~\citep{Pesquet2014,Repetti2015eusipco} at the expense of an increased number of iterations for the algorithm to converge. This feature, which has been specifically investigated for wide-band imaging \citep{Abdulaziz2017} and facet-based monochromatic imaging \citep{Naghibzedeh2018}, is not leveraged in the implementation of Algorithm~\ref{alg:pdfb} used in the experiments reported in Sections~\ref{sec:exp_simu}.

\begin{figure}
    \centering
    \begin{subfigure}{0.22\textwidth}
        \centering
        \includegraphics[width=\textwidth,keepaspectratio]{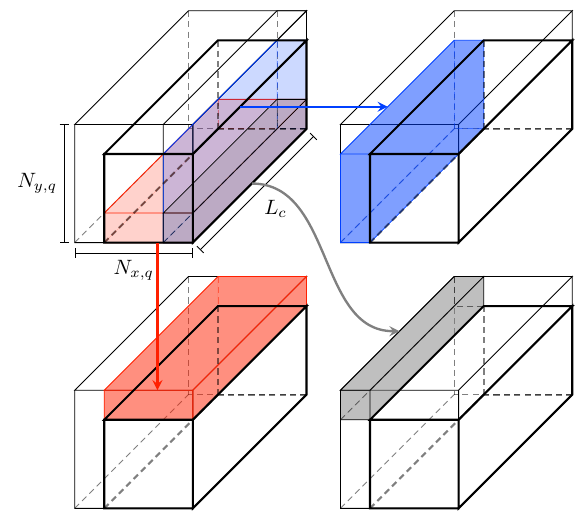}
        \caption{Broadcast values of the tile before facet update in dual space}
        \label{fig:update_gc}
    \end{subfigure}
    \qquad
    \begin{subfigure}{0.22\textwidth}
        \centering
        \includegraphics[width=\textwidth,keepaspectratio]{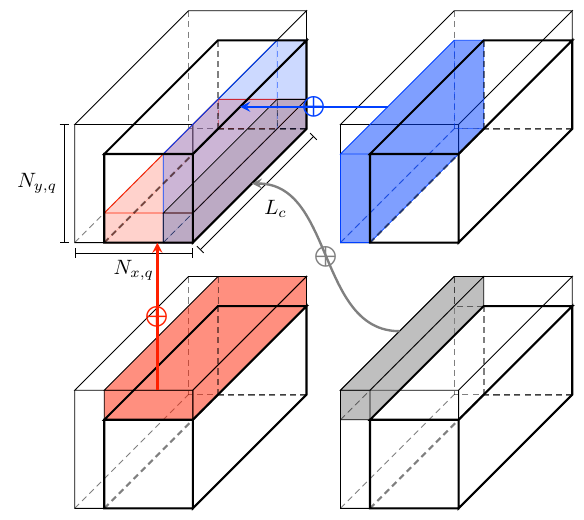}
        \caption{Broadcast and average borders before tile update in primal space
        }
        \label{fig:aggregate_gc}
    \end{subfigure}
    \caption{Illustration of the communication steps involving a facet core (represented by the top-left rectangle in each sub-figure) and a maximum of three of its neighbours. The tile underpinning each facet, located in its bottom-right corner, is delineated in thick black lines. At each iteration, the following two steps are performed sequentially. (a) Facet borders need to be completed before each facet is updated independently in the dual space (Algorithm~\ref{alg:pdfb} lines~\ref{alg:step:start_facet-nuclear}--\ref{alg:step:end_facet-l21}): values of the tile of each facet (top left) are broadcast to cores handling the neighbouring facets in order to update their borders (Algorithm~\ref{alg:pdfb} line~\ref{alg:eq:broadcast_facet}). (b) Parts of the facet tiles overlapping with borders of nearby facets need to be updated before each tile is updated independently in the primal space (Algorithm~\ref{alg:pdfb} line~\ref{alg:step:aggregate}): values of the parts of the borders overlapping with the tile of each facet are broadcast by the cores handling neighbouring facets, and averaged.
    }
    \label{fig:facet_communications}
\end{figure}

%--------------------------------------------
\subsection{Implementation, memory and computing cost} \label{ssec:cost}
To solve a spectral sub-problem $c \in \{1, \dotsc , \ncube\}$, different parallelization strategies can be adopted, depending on the computing resources available and the size of the problem to be addressed. We propose to divide the variables to be estimated among the two groups of computing cores.

\begin{itemize}
    \item \textit{Data cores:} Each core involved in this group is responsible for the update of multiple dual variables $\mathbf{v}_{c,l,b} \in \mathbb{C}^{\ndata_{c,l,b}}$ associated with the data-fidelity terms (see Algorithm~\ref{alg:pdfb} line~\ref{alg:step:dual_data1}). Each sub-problem $c$ involves $\nband_c B$ such dual variables. These cores produce auxiliary variables $\widetilde{\mathbf{v}}_{c,l,b} \in \mathbb{R}^\nbpix$, each assumed to be held in the memory of a single core (line~\ref{alg:step:dual_data2}). Note that the Fourier transform computed for each channel $l$ in line~\ref{alg:eq:fourier} is performed once per iteration on the data core $(c,l,1)$. Each data core $(c,l,b)$, with $b \in \{2, \dotsc, \nblock \}$, receives only a few coefficients of the Fourier transform of $\mathbf{x}_l$ from the data core $(c,l,1)$, selected by the operator $\mathbf{\ndata}_{c,l,b}$ (line~\ref{alg:eq:broadcast_fourier}).

    The memory requirement for each data core is $\mathcal{O}(\ndata_{c,l,b}S)$ complex numbers, dominated by the storage of the sparse {de-gridding} matrices $\mathbf{G}_{c,l,b} \in \mathbb{C}^{\ndata_{c,l,b} \times \nfourier}$ (containing $\ndata_{c,l,b}S$ nonzero values). The computing cost associated with Algorithm~\eqref{alg:pdfb} lines \ref{alg:eq:fourier}--\ref{alg:step:residual} is $\mathcal{O}(S \ndata_{c,l,b})$. Note that the computing cost on the data core $(c,l,1)$ is $\mathcal{O}(S \ndata_{c,l,1} + \nbpix \log \nbpix)$, due to the computation of the discrete Fourier transform and its adjoint (see Algorithm~\eqref{alg:pdfb}, lines~\ref{alg:eq:fourier} and \ref{alg:step:aggregate});
    \item \textit{Facet cores:} Each worker involved in this group, composed of $\nfacet$ cores, is responsible for the update of an image tile (\emph{i.e.} a portion of the primal variable) and the dual variables $\mathbf{P}_{c,q}$ and $\mathbf{W}_{c,q}$ associated with the low-rankness and the joint average sparsity prior respectively (Algorithm~\ref{alg:pdfb}, lines~\ref{alg:step:dual_lr} and~\ref{alg:step:dual_sparsity}). Note that the image cube is stored across different facet cores, which are responsible for updating an image tile (line~\ref{alg:step:primal}). Since the facets underlying the proposed prior overlap, communications involving a maximum of 4 contiguous facet cores are needed to build the facet borders prior to updating the facets independently in the dual space (Algorithm~\ref{alg:pdfb} lines~\ref{alg:step:start_facet-nuclear}--\ref{alg:step:end_facet-l21}). Values of the tile of each facet are broadcast to cores handling neighbouring facets in order to update their borders (Algorithm~\ref{alg:pdfb} line~\ref{alg:eq:broadcast_facet}, see Figure~\ref{fig:update_gc}). In a second step, parts of the facet tiles overlapping with borders of nearby facets need to be updated before each tile is updated independently in the primal space (Algorithm~\ref{alg:pdfb} line~\ref{alg:step:aggregate}). More precisely, values of the parts of the borders overlapping with the tile of each facet are broadcast by the workers handling neighbouring facets, and then averaged (see Figure~\ref{fig:aggregate_gc}). 
    From Algorithm~\eqref{alg:pdfb} lines \ref{alg:step:start_facet-nuclear}--\ref{alg:step:end_facet-l21} and \ref{alg:step:aggregate}--\ref{alg:step:primal}, the memory requirement for a facet core is $\mathcal{O}(\nband_c \max\{\widetilde{\nbpix}_q, \nbpix_q\})$ real numbers. The cost of the corresponding steps is dominated by the $9 \times \nband_c$ wavelet transforms of a matrix containing $\nbpix_q$ pixels, with asymptotic cost $\mathcal{O}(\nbpix_q)$, and the SVD of a matrix of size $\widetilde{\nbpix}_q\times \nband_c$ with an asymptotic cost $\mathcal{O}( \widetilde{\nbpix}_q\nband_c^2 + \nband_c^3)$~\cite[Table 5.5.1 p.293]{Golub2013}.
\end{itemize}

\section{Validation on synthetic data} \label{sec:exp_simu}
 
In this section, Faceted HyperSARA is validated using 
realistic simulations of wide-band radio observations of the radio galaxy Cyg~A with the Very Large Array (VLA). A first set of experiments is performed to evaluate the impact of spatial faceting, in terms of reconstruction quality and computing time, for a varying number of facets $\nfacet$ and a varying size of the overlapping regions, on a single spectral sub-problem ($\ncube=1$). A second set of experiments is performed to assess the impact of spectral faceting, for a varying number of spectral sub-problem $\ncube$, using a single underlying facet along the spatial dimension ($\nfacet=1$). Simulation results are systematically compared with those of both SARA and HyperSARA. All algorithms have been implemented in MATLAB.

%--------------------------------------------
\subsection{Simulation setting}
\label{sec:sim_setting}

Firstly, the original Cyg~A image cube, composed of $\nband$ spectral channels, is simulated from a reference spectral channel image using a power-law spectral model~\citep{Rau2011}. All simulations rely on a recently published map of Cyg~A at S band (2~GHz)~\citep{Dabbech2021}.
Considering the frequencies $(\nu_l)_{1 \leq l\leq \nband}$, the spectral model reads, for $n \in\{1, \dotsc, \nbpix\}$ and $l \in \{2, \dotsc, \nband \}$ 
\begin{equation}
  x_{n, l} = x_{n, 1} \big( \frac{\nu_l}{\nu_1}\big)^{ {\check{x}_{a, n}}+ {\check{x}_{b, n}} \log(\nu_l/\nu_1)},
\end{equation}
where $\mathbf{x}_1 = (x_{n, 1})_{1 \leq n \leq \nbpix} \in \mathbb{R}^\nbpix$ is the reference image at the frequency $\nu_1$, $\check{\mathbf{x}}_a = (\check{x}_{a, n})_{1 \leq n \leq \nbpix} \in \mathbb{R}^\nbpix$ and $\check{\mathbf{x}}_b = (\check{x}_{b, n})_{1 \leq n \leq \nbpix} \in \mathbb{R}^\nbpix$ are the spectral-index and spectral-curvature maps, respectively. On the one hand, the spectral-curvature map $\check{\mathbf{x}}_{b}$ is simulated as a random Gaussian field \citep{Kroese2015}. On the other hand, the spectral-index map $\check{\mathbf{x}}_{a}$ is obtained from the reference spectral channel image ${\mathbf{x}}_1$ to ensure its spatial correlation \citep{Junklewitz2015,Abdulaziz2016}. More precisely, the spectral map is simulated as the sum of two components: the logarithm of a smoothed version of $\mathbf{x}_1$ and a random Gaussian field. 

Secondly, the wide-band measurements are simulated in absence of DDEs using configurations A and C of the VLA, with a total observation time of about 7~hours and 11~hours, respectively and an integration time of 10 seconds. The de-gridding matrix $\mathbf{G}$ considered relies on Kaiser-Bessel kernels of size $S = 7 \times 7$ \citep{Fessler2003}. The frequency channels are uniformly sampled, such that the associated $uv$-coverage is a scaled version of the $uv$-coverage at the reference frequency $\nu_1$ by the ratio $\nu_l/\nu_1$, for $1<l \leq \nband$. The visibilities are corrupted by an additive, zero-mean complex white Gaussian noise. An input signal-to-noise ratio (iSNR) of 40 dB is considered, which is defined as 
\begin{linenomath}
\begin{equation}
  \displaystyle \iSNR = 10 \log_{10}\left( \sum_{l=1}^\nband \frac{\norm{\bs{\Phi}_l\overline{\mathbf{x}}_l}^2_2}{\ndata_l \sigma^2_l} \right).
\end{equation}
\end{linenomath}
Note that, given the larger computational cost of HyperSARA, the number of visibilities, the image size and the number of channels are chosen so that the algorithms can be run in a reasonable amount of time for the different simulation scenarios described below, while highlighting the respective interest of spatial and spectral faceting.

Thirdly, we comment on the chosen values for algorithmic parameters of interest, common to both the spatial faceting and the spectral faceting experiments. All the experiments rely on $\nblock=1$ data block per channel, as the moderate data size considered does not require more. The regularization parameters have all been set using the heuristics described in Sections~\ref{Ssec:SARA}, \ref{Ssec:hyperSARA} and~\ref{sec:faceting_prior}, for SARA, HyperSARA, and Faceted HyperSARA respectively. The following values have been considered for the stopping criteria involved in Algorithms~\ref{alg:reweighting} and~\ref{alg:pdfb}, and also used in SARA and HyperSARA: $\nreweightmax = 5$, $\npdfbmin = 10$, $\npdfbmax = 2000$, $\relvarrw = \relvarpdfb = 10^{-5}$.

Finally, the Sections~\ref{ssec:exp:spatial_faceting} and~\ref{ssec:exp:spectral_faceting} below describe the specificities of the spatial and spectral faceting experiments respectively.

\subsubsection{Spatial faceting} \label{ssec:exp:spatial_faceting}

The performance of Faceted HyperSARA is evaluated for different parameters of the spatial faceting. Wide-band data are generated utilizing the $uv$-coverage obtained by the A configuration of the VLA, and a reference spectral channel image of size $\nbpix = 1024 \times 2048$ with pixel size $\delta\ell=0.0683~\textrm{arcsec}$, that is a re-sized version of the original image provided in~\citet{Dabbech2021}. A number of $\nband = 20$ spectral channels is considered, uniformly sampled in the frequency range $[2.052,3.572]$~GHz. The number of measurements per channel is $\ndata_l = 7.62 \times 10^5$, leading to image cube and data close to 1 Gigabyte in size. To guarantee the low-rankness prior plays a relevant role, each optimization sub-problem~\eqref{eq:problem-split-subcube} should involve a reasonable number of channels. Given the number of spectral channels considered, all spatial faceting experiments are thus conducted with $\ncube=1$ spectral facets (\emph{i.e.} without spectral faceting). Performance assessment is conducted for (i) a varying number of facets $\nfacet$ along the spatial dimensions and a fixed spatial overlap for the low-rankness regularization; (ii) a fixed number of facets and a varying spatial overlap. Additional details can be found in the following lines. 
\begin{itemize}
\item \textit{Varying overlap:} Reconstruction performance and computing time are evaluated with $\nfacet = 16$ (4 facets along each spatial dimension) and a varying size of the overlapping region for the faceted nuclear norm (0\%, 10\%, 25\%, 40\% and 50\% of the spatial size of the facet) in each of the two spatial dimensions. Note that the overlap involved in the average joint-sparsity prior is fixed by design of the faceted wavelet transform~\citep{Prusa2012}. 
\item \textit{Varying number of facets:} The reconstruction performance and computing time of Faceted HyperSARA are reported for experiments with $\nfacet \in \{4, 9, 16\}$ (corresponding to 2, 3 and 4 facets along each spatial dimension) with a fixed overlap for the faceted low-rankness prior corresponding to 10\% of the spatial size of a facet.
\end{itemize}

\subsubsection{Spectral faceting} \label{ssec:exp:spectral_faceting}
The effect of the spectral faceting is evaluated in terms of computing time and reconstruction quality using wide-band data generated with the $uv$-coverage given by the C configuration of the VLA, from a reference spectral channel image of size $\nbpix = 256 \times 512$ with a pixel size $\delta\ell=0.8~\textrm{arcsec}$, that is a down-sized version of the original image provided in \citet{Dabbech2021}. A number of $\nband = 100$ frequency channels is considered, uniformly sampled in the frequency range $[2.052,3.636]$~GHz. The number of measurements per channel is $\ndata_l = 2.68 \times 10^5$, leading to image cube and data size close to 1 Gigabyte in size. Note that the spatial faceting experiments from Section~\ref{ssec:exp:spatial_faceting} already assess whether the image quality obtained with HyperSARA can be preserved by Faceted HyperSARA using $\nfacet > 1$ spatial facets. Without loss of generality and given the small image size considered for this experiment, the overall reconstruction performance of Faceted HyperSARA is thus evaluated with a single facet along the spatial dimension ($\nfacet = 1$). A varying number of sub-problems $\ncube\in\{2,5,10\}$ is considered (indexed by $c \in \{1, \dotsc, \ncube\}$), with $\nband_c=\nband/\ncube$ spectral facets. The results are compared with those of HyperSARA (\emph{i.e.} Faceted HyperSARA with $\nfacet=1$ and $\ncube=1$) and SARA (\emph{i.e.} Faceted HyperSARA with $\nfacet=1$, $\ncube=\nband$ and $\overline{\mu}_{c, q} = 0$).

\begin{table*}
    \centering
    \resizebox{0.98\textwidth}{!}{%
    \begin{tabular}{lrrrrrrrr} \toprule
        & $\aSNR$ ($\pm \sSNR$) (dB)  & $\aSNR_{\log}$ ($\pm \sSNR_{\log}$) (dB) & CPU cores & PDFB iterations & $\runp$ (s) & $\run$ (h) & $\cpup$ (s) & $\cpu$ (h) \\ \midrule
        SARA
        & \textbf{35.05 ($\pm 5.90e-01$)} & 13.72 ($\pm 1.59e-01$) & 240 & 3275 & \textbf{3.28 ($\pm 3.79e-01$)} & 3.38 & \textbf{7.13 ($\pm 9.50e-01$)} & 129.77 \\
        HyperSARA
        & \textbf{39.47 ($\pm  2.15e+00$)} & 16.36 ($\pm 1.27e+00$) & 22  & 9236 & \textbf{25.36 ($\pm 8.51e-01$)} & 65.06 & \textbf{84.49 ($\pm 2.79e+00$)} & 216.76 \\
        Faceted HyperSARA (0\% overlap)
        & \textbf{40.03 ($\pm 2.41e+00$)} & 16.79 ($\pm 1.42e+00$) & 36 & 10961 & \textbf{11.55 ($\pm 7.03e-01$)} & 35.18 & \textbf{284.17 ($\pm 1.34e+01$)} & 865.22 \\
        Faceted HyperSARA (10\% overlap)
        & \textbf{ 40.08 ($\pm 2.40e+00$)} & 16.85 ($\pm 1.43e+00$) & 36 & 10945 & \textbf{11.62 ($\pm 4.97e-01$)} & 35.32 & \textbf{286.06 ($\pm 1.08e+01$)} & 869.71  \\
        Faceted HyperSARA (25\% overlap)
        & \textbf{40.22 ($\pm 2.41e+00$)} & 16.96 ($\pm 1.47e+00$) & 36 & 10918 &  \textbf{11.96 ($\pm 5.34e-01$)} & 36.26 & \textbf{290.71 ($\pm 1.39e+01$)} & 881.66 \\
        Faceted HyperSARA (40\% overlap)
        & \textbf{40.24 ($\pm 2.42e+00$)} & 17.05 ($\pm 1.51e+00$) & 36 & 10934 & \textbf{12.67 ($\pm 5.45e-01$)} & 38.47 & \textbf{298.32 ($\pm 1.43e+01$)} & 906.08 \\ 
        Faceted HyperSARA (50\% overlap)
        & \textbf{40.08 ($\pm 2.53e+00$)} & 16.95 ($\pm 1.52e+00$) & 36 & 10962 & \textbf{13.69 ($\pm 6.48e-01$)} & 41.68 & \textbf{311.14 ($\pm 1.60e+01$)} & 947.41 \\ 
        \bottomrule 
    \end{tabular}
    }
    \caption{Spatial faceting experiment: varying size of the overlap region for the faceted nuclear norm regularization. Reconstruction performance of Faceted HyperSARA with $(\nfacet, \ncube)=(16, 1)$, compared to HyperSARA (\emph{i.e.} Faceted HyperSARA with $(\nfacet, \ncube)=(1,1)$) and SARA (\emph{i.e.} Faceted HyperSARA with $(\nfacet, \ncube, (\overline{\mu}_{c, q})_{c,q})=(1, \nband, \mathbf{0}_{\ncube \times \nfacet})$). The results are reported in terms of
    average $\aSNR$ and $\aSNR_{\log}$ (both in dB with the associated standard deviation $\sSNR$ and $\sSNR_{\log}$), runtime and CPU time per iteration per sub-cube ($\runp$ and $\cpup$ in seconds, with the associated standard deviation over the iterations and the $\ncube$ spectral subcubes). The number of CPU cores, the total number of PDFB iterations until convergence and the total runtime and CPU time to reconstruct the full image cube are also reported. The reconstruction $\SNR$ and the runtime per iteration are highlighted in bold face.}
    \label{tab:facet_overlap}
\end{table*}

%--------------------------------------------
\subsection{Hardware} 

All the methods compared in this section have been run on compute nodes of Cirrus, one of the UK's Tier2 HPC services\footnote{\url{https://epsrc.ukri.org/research/facilities/hpc/tier2/}}. Cirrus is an SGI ICE XA system composed of 280 compute nodes, each with two 2.1~GHz, 18-core, Intel Xeon E5-2695 (Broadwell) series processors (36 CPU cores in total per node). The compute nodes have 256~Gigabyte of memory shared between the two processors. The system has a single Infiniband FDR network connecting nodes with a bandwidth of 54.5~GB/s.

All algorithms have been run in the following configurations for the considered experiments, taking into account the different number of visibilities considered in the two simulation scenarios, and the extra parallelization flexibility offered by image faceting.

\begin{itemize}
  \item \textbf{SARA}.
  \begin{enumerate}
      \item[] \noindent For all the experiments considered, SARA uses 12 CPU cores to reconstruct a single channel, among which 9 CPU cores are more specifically used to handle the average sparsity terms (associated with the nine bases of the SARA dictionary), following the parallelization strategy proposed by~\citet{Onose2016}. The wideband image reconstruction has been conducted in parallel with a single node for each single channel problem, thus simultaneously leveraging $12 \times \nband$ cores across $\nband$ nodes.
  \end{enumerate}
  \item \textbf{HyperSARA}. 
  \begin{enumerate}
      \item \textit{Spatial faceting experiments}: the full problem~\eqref{eq:problem}-\eqref{eq:hypersara0} is addressed with 22 CPU cores: 20 CPU cores for both the average-sparsity and the data fidelity dual variables (1 CPU core per frequency channel), and 2 CPU cores for both the low-rankness dual variable and the primal variable. A single compute node has been used to run the algorithm.
      \item \textit{Spectral faceting experiments}: the full problem~\eqref{eq:problem}-\eqref{eq:hypersara0} is addressed with 16 CPU cores for any spectral faceting experiment: 10 CPU cores for both the average-sparsity and the data fidelity dual variables, and 6 CPU cores for both the low-rankness dual variable and the primal variable. A single compute node has been used to run the algorithm.
  \end{enumerate}
  \item \textbf{Faceted HyperSARA}:
  \begin{enumerate}
      \item \textit{Spatial faceting experiments}: to address each independent sub-problem in~\eqref{eq:problem-split-subcube}, 20 CPU cores are used for the data fidelity terms (1 CPU core per frequency channel), and 1 CPU core is taken for each of the $\nfacet$ facets (up to 36 cores in total). A single compute node has been used to run the algorithm.
      \item \textit{Spectral faceting experiments}: to address each independent sub-problem in~\eqref{eq:problem-split-subcube}, the same number of cores as HyperSARA is used for the different number of spectral sub-cube $\ncube \in \{2, 5, 10 \}$ considered. Note that the independent sub-problems~\eqref{eq:problem-split-subcube} have been solved in parallel using Algorithm~\ref{alg:reweighting} on a single compute node each, thus simultaneously leveraging $16 \times \ncube$ cores across $\ncube$ nodes.
  \end{enumerate}
\end{itemize}

\begin{figure*}
    \begin{center}
        \begin{subfigure}{\textwidth}
            \includegraphics[width=0.48\textwidth,keepaspectratio]{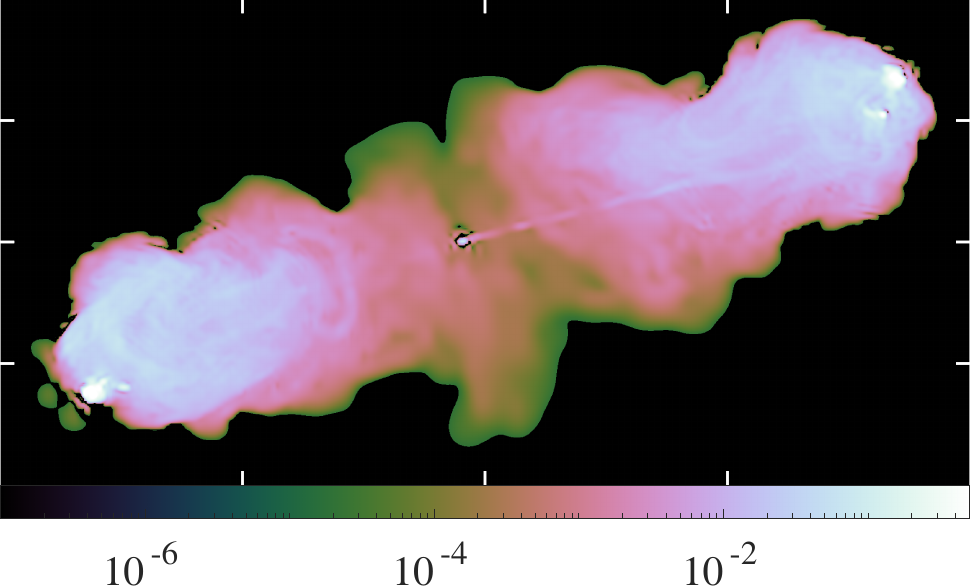}\\
            \includegraphics[width=0.48\textwidth,keepaspectratio]{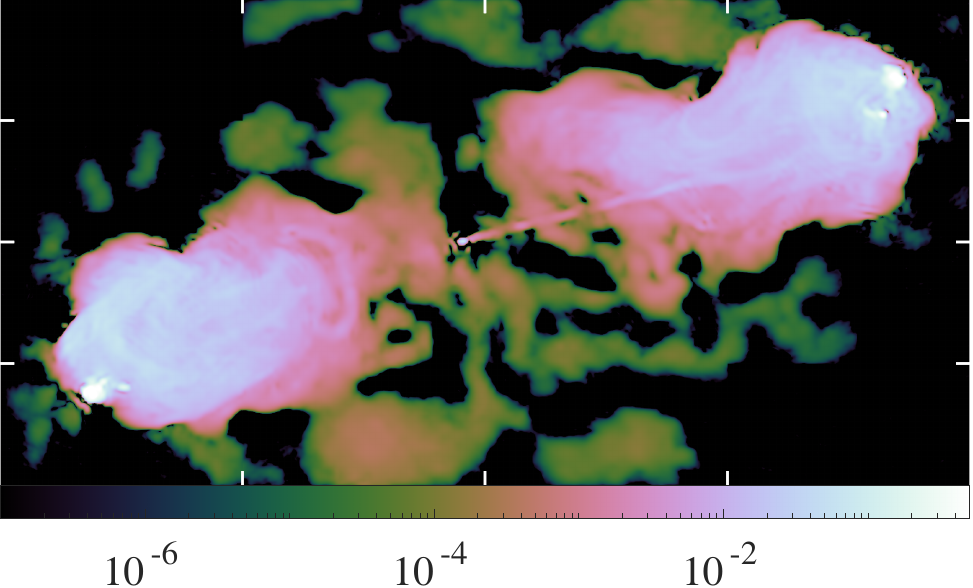}
            \raisebox{-0.45cm}{\includegraphics[width=0.486\textwidth,keepaspectratio]{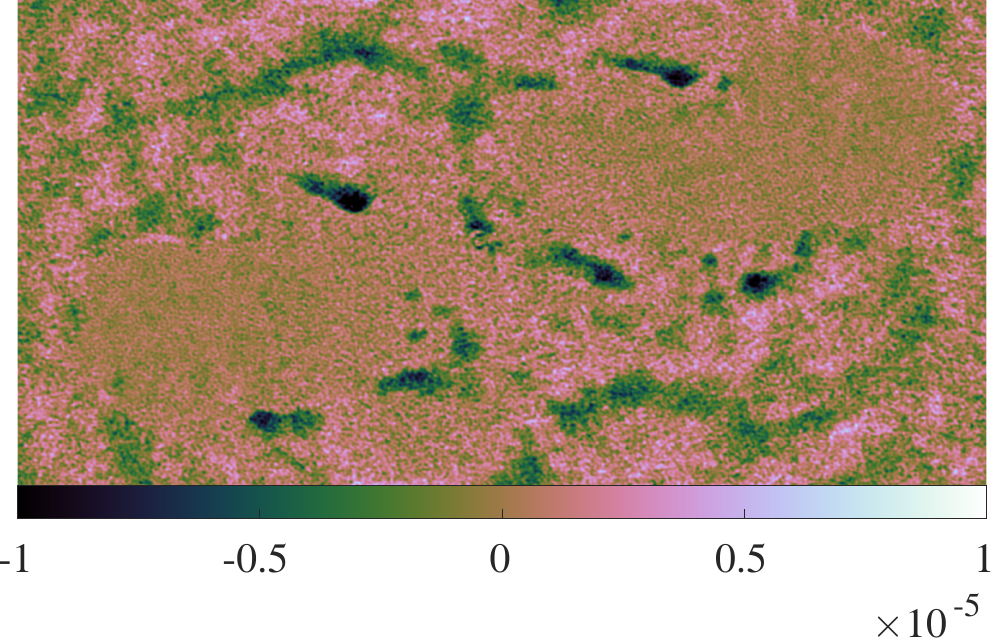}}\\
            \includegraphics[width=0.48\textwidth,keepaspectratio]{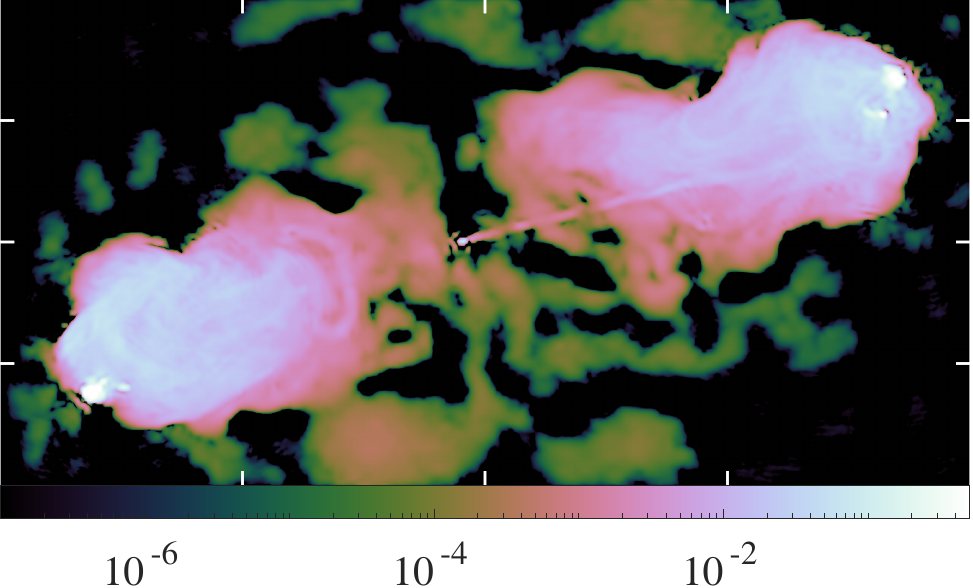}
            \raisebox{-0.45cm}{\includegraphics[width=0.486\textwidth,keepaspectratio]{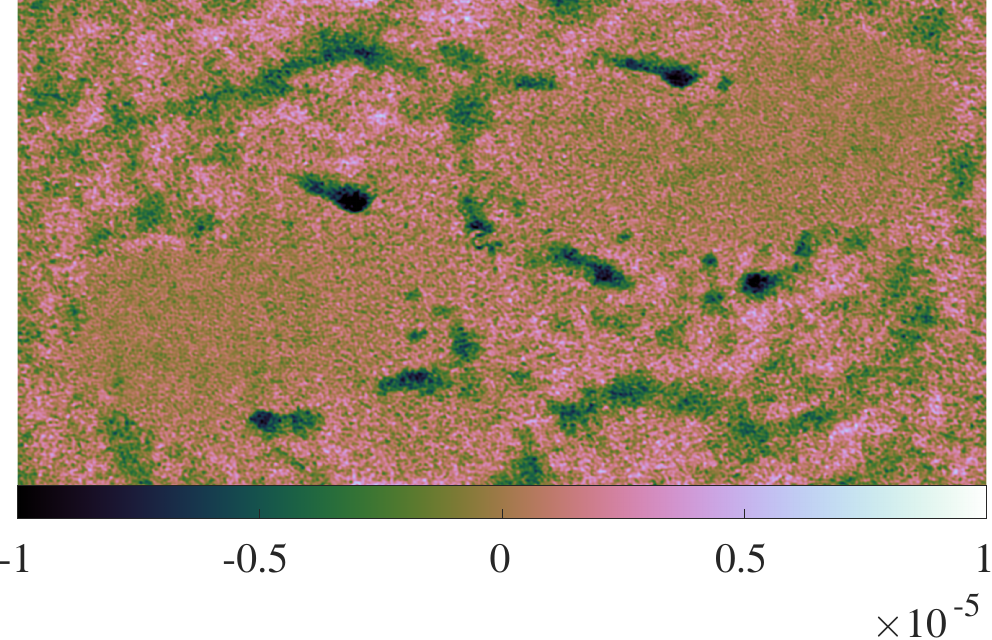}}\\
            \includegraphics[width=0.48\textwidth,keepaspectratio]{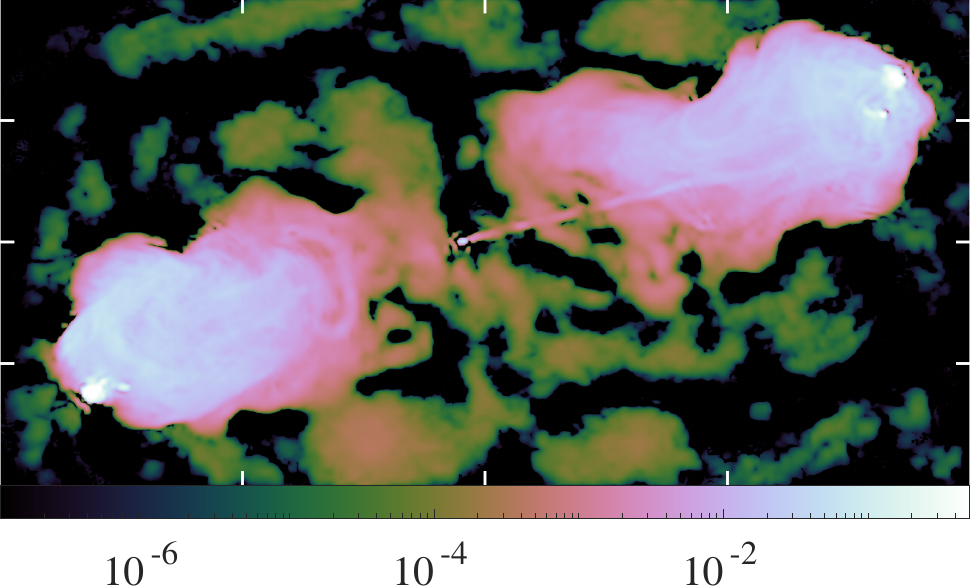}
            \raisebox{-0.45cm}{\includegraphics[width=0.486\textwidth,keepaspectratio]{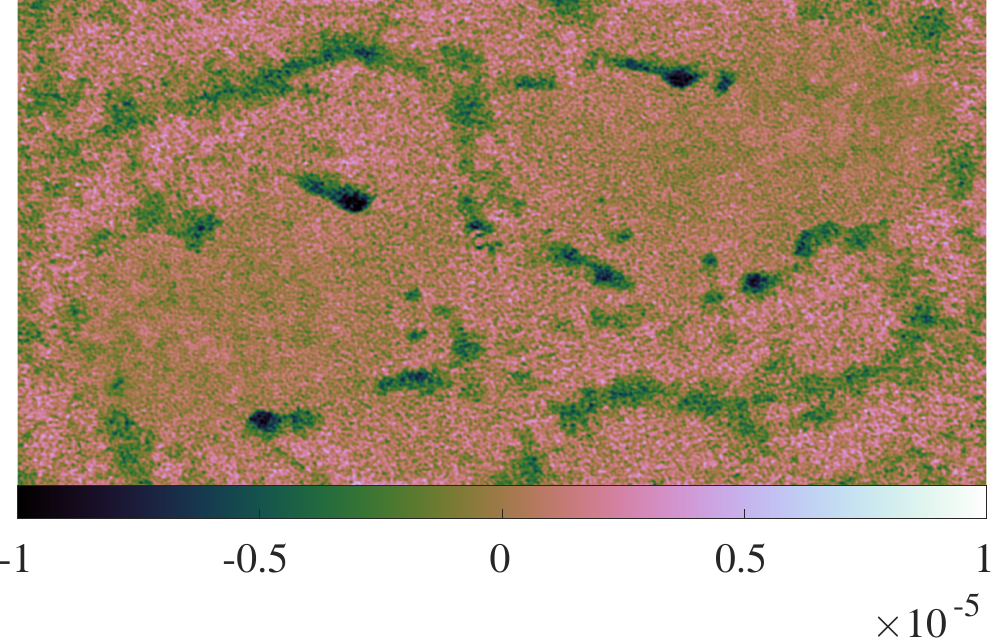}}
        \end{subfigure}
    \end{center}
    
    \vspace*{-0.2cm}
    
    \caption{Spatial faceting experiment: reconstructed images of channel $\nu_{1} = 2.052$ GHz of an image cube with $L=20$ spectral channels, obtained with Faceted HyperSARA (using $(\nfacet,\ncube)=(16,1)$) and HyperSARA (\emph{i.e.} Faceted HyperSARA with $(\nfacet,\ncube)=(1,1)$). Top row: ground truth image (Jy/pixel) displayed in $\log_{10}$ scale. In rows $2-4$ are reported the reconstructed images of Faceted HyperSARA without overlap, Faceted HyperSARA with 10\% overlap and HyperSARA, respectively. From left to right, model image (Jy/pixel) displayed in $\log_{10}$ scale and residual image normalized by $\|\bs{\Phi}_{1}\|_{\text{S}}^2$, displayed in linear scale. Horizontal and vertical marks at the panel edges identify the contour of each tile.}
    \label{fig:images_overlap_1}
\end{figure*}

\begin{figure*}
    \begin{center}
        \begin{subfigure}{\textwidth}
            \imdraw{0.48\textwidth}{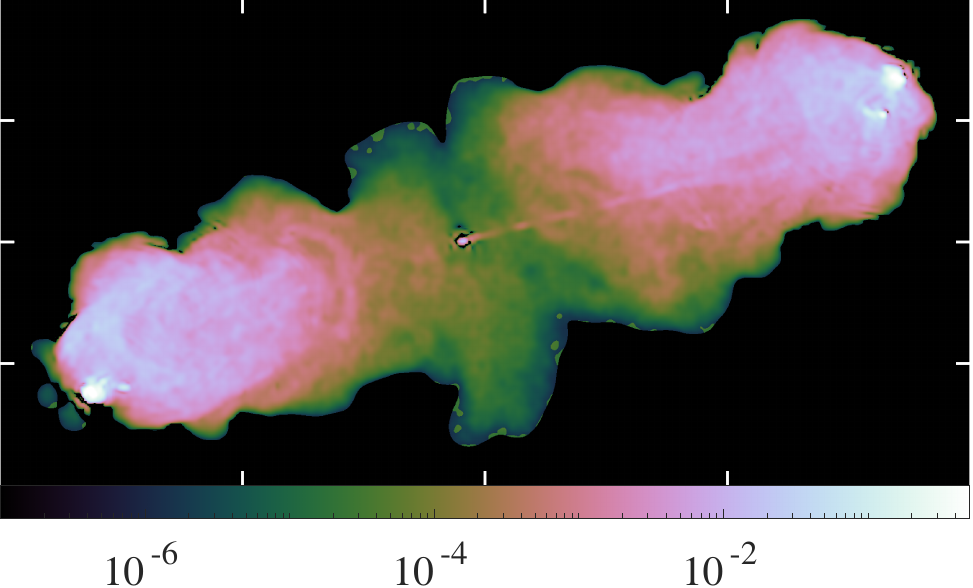}\\
            \imdraw{0.48\textwidth}{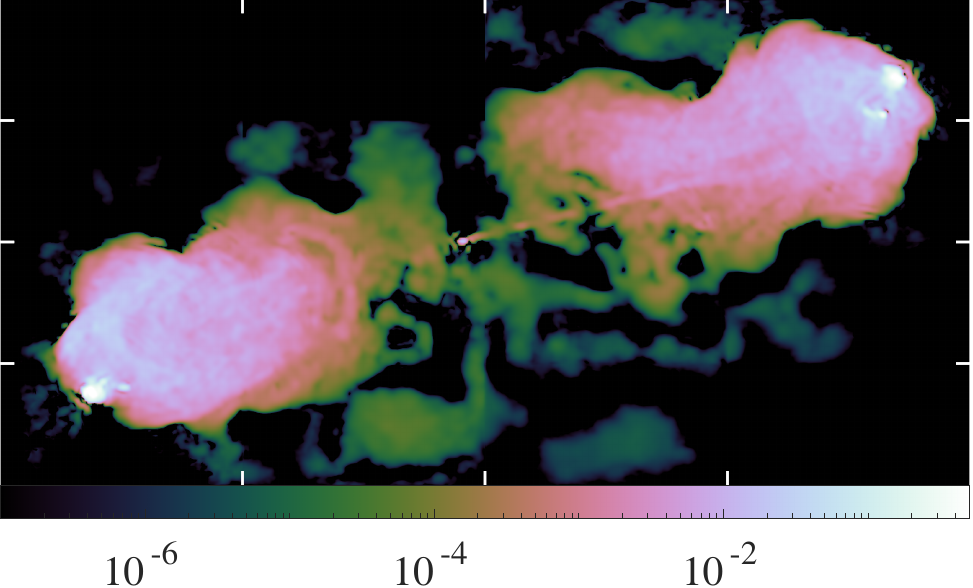}
            \raisebox{-0.45cm}{\includegraphics[width=0.486\textwidth,keepaspectratio]{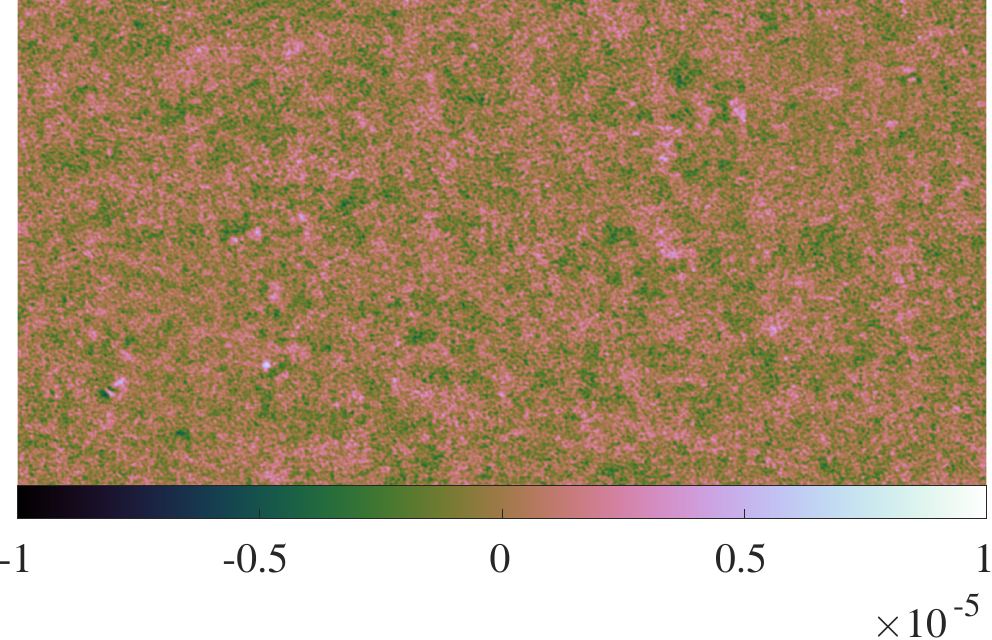}}\\
            \imdraw{0.48\textwidth}{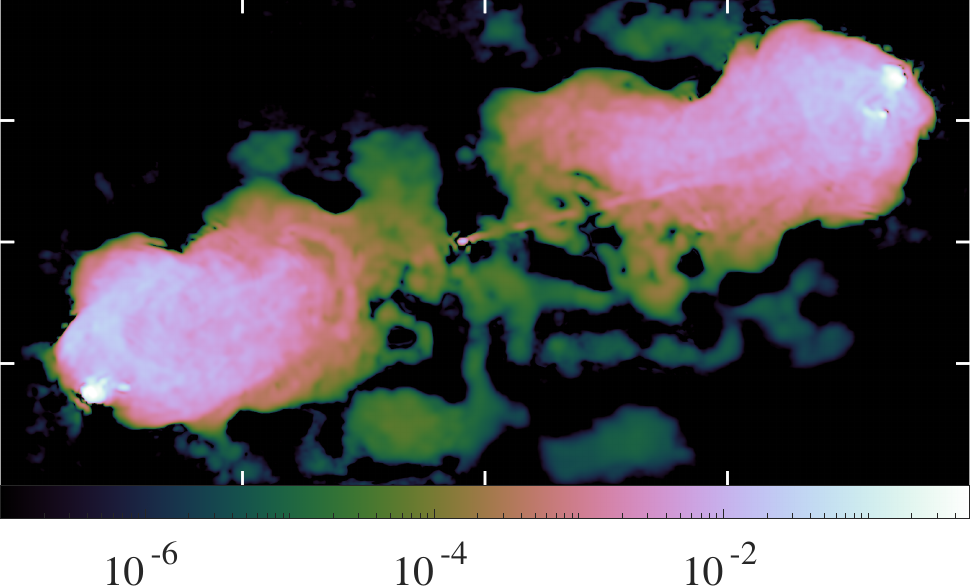}
            \raisebox{-0.45cm}{\includegraphics[width=0.486\textwidth,keepaspectratio]{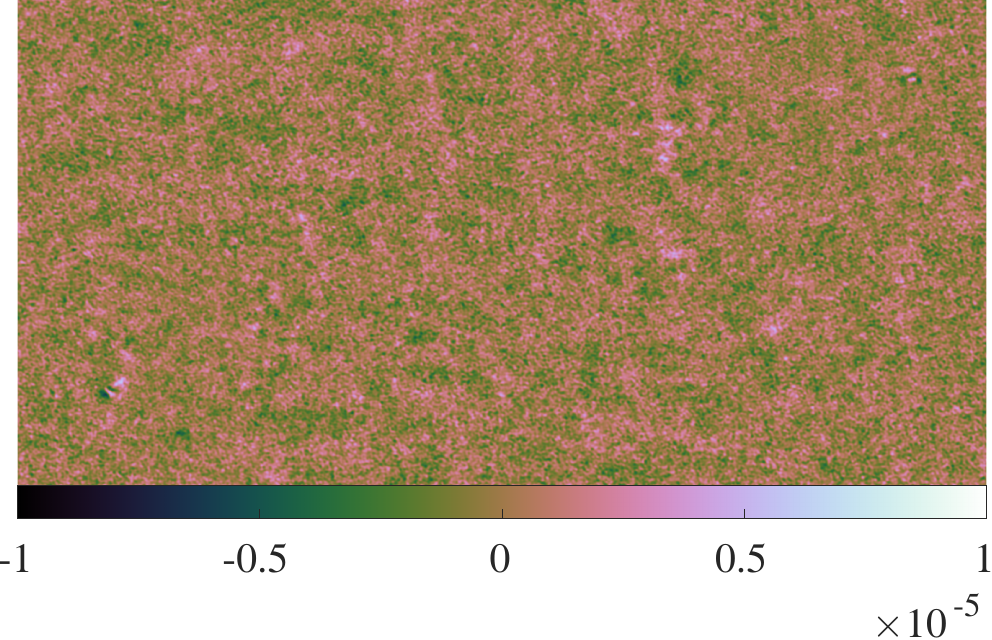}}\\
            \imdraw{0.48\textwidth}{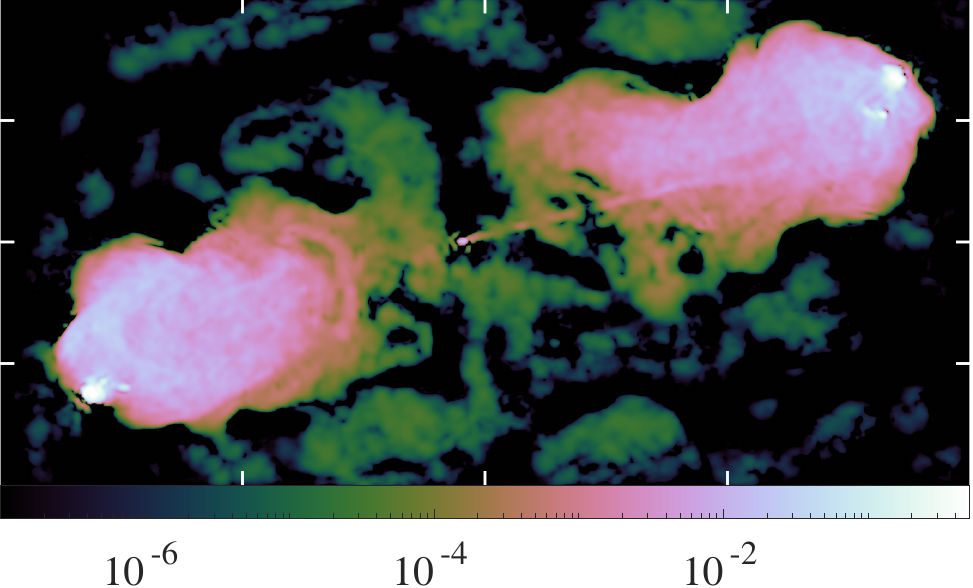}
            \raisebox{-0.45cm}{\includegraphics[width=0.486\textwidth,keepaspectratio]{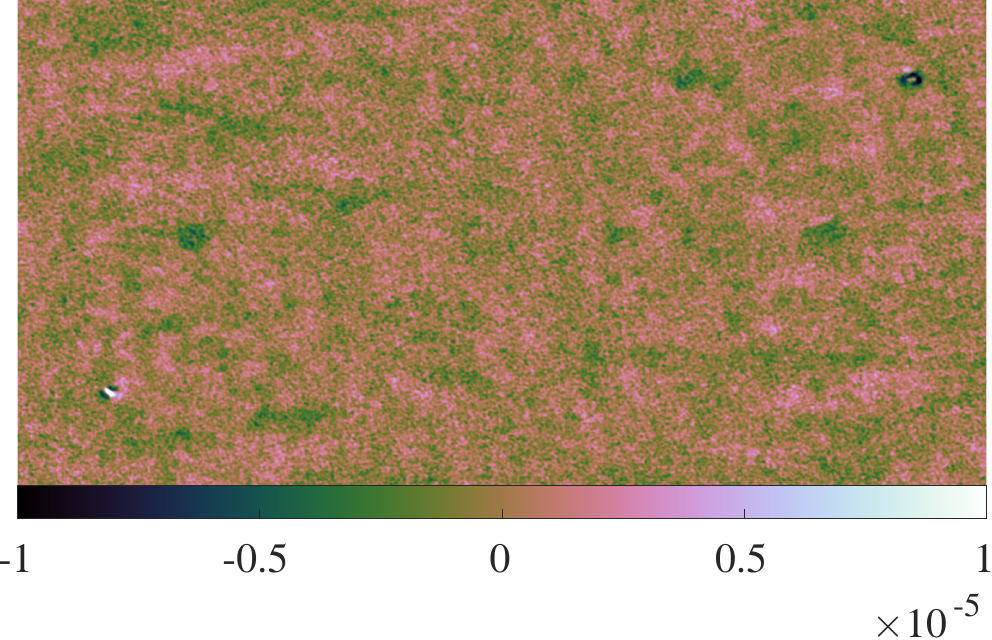}}
        \end{subfigure}
    \end{center}
    
    \vspace*{-0.2cm}
    
    \caption{Spatial faceting experiment: reconstructed images of channel $\nu_{20} = 3.572$ GHz of an image cube with $L=20$ spectral channels, obtained with Faceted HyperSARA ($(\nfacet,\ncube)=(16,1)$) and HyperSARA (\emph{i.e.} Faceted HyperSARA with $(\nfacet,\ncube)=(1,1)$). Top row: ground truth image (Jy/pixel) displayed in $\log_{10}$ scale. In rows $2-4$ are reported the reconstructed images of Faceted HyperSARA without overlap, Faceted HyperSARA with 10\% overlap, and HyperSARA, respectively. From left to right, model image (Jy/pixel) displayed in $\log_{10}$ scale and residual image normalized by $\|\bs{\Phi}_{20}\|_{\text{S}}^2$, displayed in linear scale. Horizontal and vertical marks at the panel edges identify the contour of each tile. In absence of overlap, and in this case only, facet artefacts emerge in the model image obtained with Faceted HyperSARA. Regions where artefacts appear in absence of facet overlap are delineated in white in all the figures to ease the comparison.}
    \label{fig:images_overlap_2}
\end{figure*}

\begin{table*}
    \centering
    \resizebox{0.98\textwidth}{!}{%
    \begin{tabular}{lrrrrrrrr} \toprule
        & $\aSNR$ ($\pm \sSNR$) (dB)  & $\aSNR_{\log}$ ($\pm \sSNR_{\log}$) (dB) & CPU cores & PDFB iterations & $\runp$ (s) & $\run$ (h) & $\cpup$ (s) & $\cpu$ (h) \\ \midrule
        SARA
        & \textbf{35.05 ($\pm 5.90e-01$)} & 13.72 ($\pm 1.59e-01$) & 240 & 3275 & \textbf{3.28 ($\pm 3.79e-01$)} & 3.38 & \textbf{7.13 ($\pm 9.50e-01$)} & 129.77 \\
        HyperSARA
        & \textbf{39.47 ($\pm  2.15e+00$)} & 16.36 ($\pm 1.27e+00$) & 22  & 9236 & \textbf{25.36 ($\pm 8.51e-01$)} & 65.06 & \textbf{84.49 ($\pm 2.79e+00$)} & 216.76 \\
        Faceted HyperSARA ($\nfacet=4$)
        & \textbf{39.79 ($\pm 2.34e+00$)} & 16.47 ($\pm 1.39e+00$) & 24 & 10989  & \textbf{26.50 ($\pm 1.88e+00$)} & 80.90 & \textbf{184.41 ($\pm 9.22e+00$)} & 562.90 \\
        Faceted HyperSARA ($\nfacet=9$)
        & \textbf{40.00 ($\pm 2.40e+00$)} & 16.82 ($\pm 1.42e+00$) & 29 & 11009 & \textbf{15.38 ($\pm 1.38e+00$)} & 47.04 & \textbf{226.52($\pm 1.10e+01$)} & 692.71 \\
        Faceted HyperSARA ($\nfacet=16$)
        & \textbf{ 40.08 ($\pm 2.40e+00$)} & 16.85 ($\pm 1.43e+00$) & 36 & 10945 & \textbf{11.62 ($\pm 4.97e-01$)} & 35.32 & \textbf{286.06 ($\pm 1.08e+01$)} & 869.71  \\
        \bottomrule
    \end{tabular}
    }
    \caption{Spatial faceting experiment: varying number of facets along the spatial dimension $\nfacet$. Reconstruction performance of Faceted HyperSARA ($\ncube=1$, overlap of 10\%), compared to HyperSARA (\emph{i.e.} Faceted HyperSARA with $(\nfacet,\ncube)=(1, 1)$) and SARA (\emph{i.e.} Faceted HyperSARA with $(\nfacet, \ncube, (\overline{\mu}_{c, q})_{c,q})=(1, \nband, \mathbf{0}_{\ncube \times \nfacet})$). The results are reported in terms of
    average $\aSNR$ and $\aSNR_{\log}$ (both in dB, with the associated standard deviation over frequencies $\sSNR$ and $\sSNR_{\log}$), runtime and CPU time per iteration per sub-cube ($\runp$ and $\cpup$ in seconds, with the associated standard deviation over the iterations and the $C$ spectral subcubes). The number of CPU cores, the total number of PDFB iterations until convergence and the total runtime and CPU time to reconstruct the full image cube are also reported. The reconstruction $\SNR$ and the runtime per iteration are highlighted in bold face.} 
    \label{tab:facet_number}
\end{table*}

%--------------------------------------------
\subsection{Evaluation metrics} \label{ssec:metrics}

Imaging performance is evaluated in terms of the reconstruction $\SNR$, defined for each channel $l \in \{1, \dotsc, \nband \}$ as
\begin{linenomath}
\begin{equation}
	\SNR_l(\mathbf{x}_l) = 20 \log_{10} \left( \frac{\norm{\overline{\mathbf{x}}_l}_2}{\norm{\overline{\mathbf{x}}_l - \mathbf{x}_l}_2} \right).
\end{equation}
\end{linenomath}
Results are reported in terms of the average and standard deviation of $\SNR_l(\mathbf{x}_l)$ over the spectral channels $l$, abbreviated as $\aSNR$ and $\sSNR$, respectively.
However, the above criterion shows limitations to reflect the dynamic range of the estimated image, and thus appreciate improvements in the reconstruction quality of faint emissions. As a complementary information, the following criterion is computed for each channel in $\log_{10}$ scale

\begin{linenomath}
\begin{multline}
\SNR_{\log, l} (\mathbf{x}_l) = \\
    20 \log_{10}\left( \frac{\norm{\log_{10}(\overline{\mathbf{x}}_l/\epsilon_l+  \mathbf{1}_\nbpix)}_2}{\norm{\log_{10}(\overline{\mathbf{x}}_l/\epsilon_l + \mathbf{1}_\nbpix) - \log_{10}( \mathbf{x}_l/\epsilon_l + \mathbf{1}_\nbpix)}_2} \right),
\end{multline}
\end{linenomath}
where the $\log_{10}$ function is applied term-wise, and $\epsilon_l$ is set to the noise level estimate in image space defined as $\sigma_l/\norm{\bs{\Phi}_l}_\text{S}$. Results are reported in terms of the average and standard deviation of the $\SNR_{\log,l}(\mathbf{x}_l)$ over the spectral channels $l$, abbreviated as $\aSNR_{\log}$ and $\sSNR_{\log}$, respectively.
The $\aSNR_{\log}$ criterion provides a more balanced information between the reconstruction quality of the brightest and the faintest emissions across all spectral channels, for a given dynamic range set by the choice of $\epsilon_l$.

The computational complexity of the algorithms is assessed using the following criteria:
\begin{enumerate}
  \item runtime per PDFB iteration per sub-cube ($\runp$), averaged over all PDFB iterations and sub-cubes, and given with the associated standard deviation. Note that the runtime is affected by the efficiency of the parallel implementation considered and potential load balancing issues;
  \item (active) CPU time per PDFB iteration per sub-cube ($\cpup$), averaged over all PDFB iterations and sub-cubes. Figures are given with the associated standard deviation, omitting the communication time between workers. This criterion gives more insight into the computational cost incurred by the different approaches on a per iteration basis, as it is not affected by the parallelization scheme adopted;
  \item total runtime until convergence ($\run$);
  \item total (active) CPU time used to reconstruct the full image cube ($\cpu$).
\end{enumerate}

\begin{figure*}
    \begin{center}
        \begin{subfigure}{\textwidth}
            \includegraphics[width=0.48\textwidth,keepaspectratio]{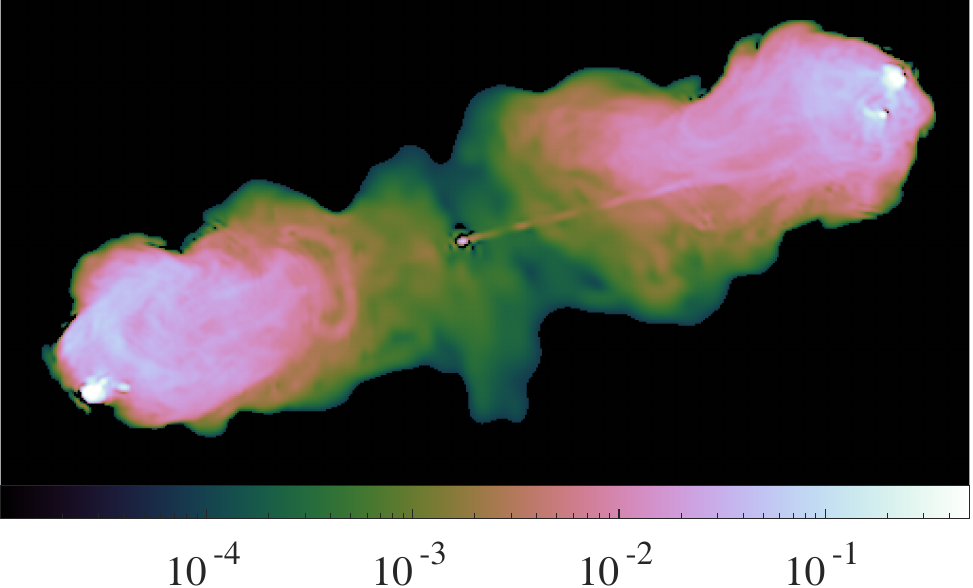}\\
            \includegraphics[width=0.48\textwidth,keepaspectratio]{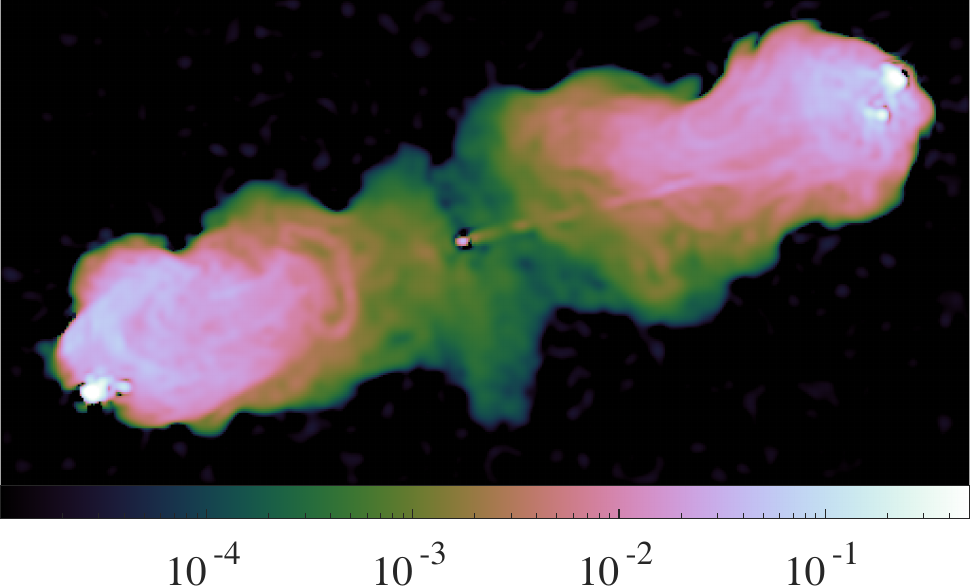}
            \raisebox{-0.45cm}{\includegraphics[width=0.49\textwidth,keepaspectratio]{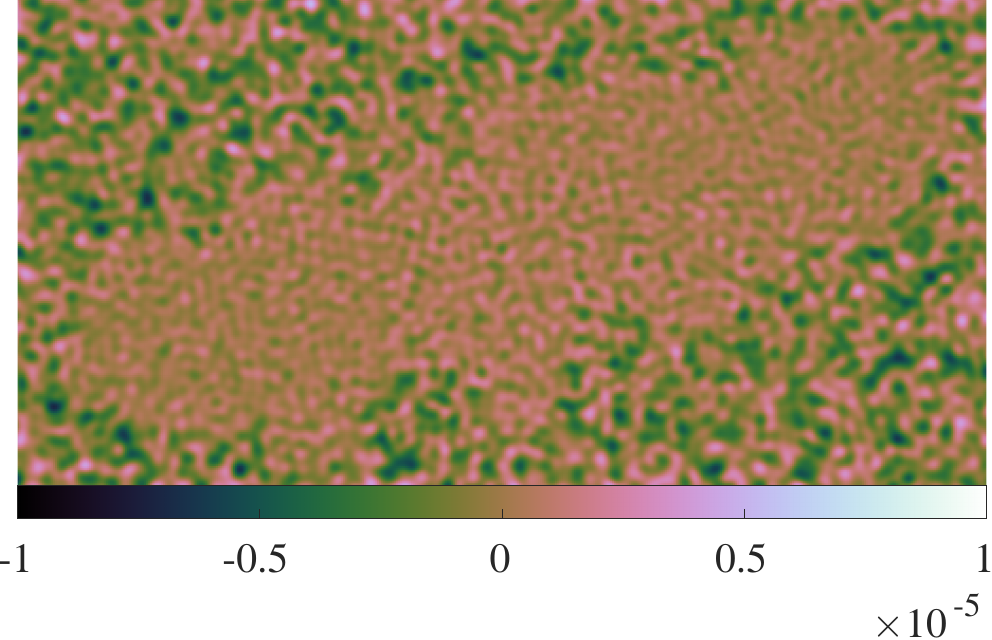}}\\
            \includegraphics[width=0.48\textwidth,keepaspectratio]{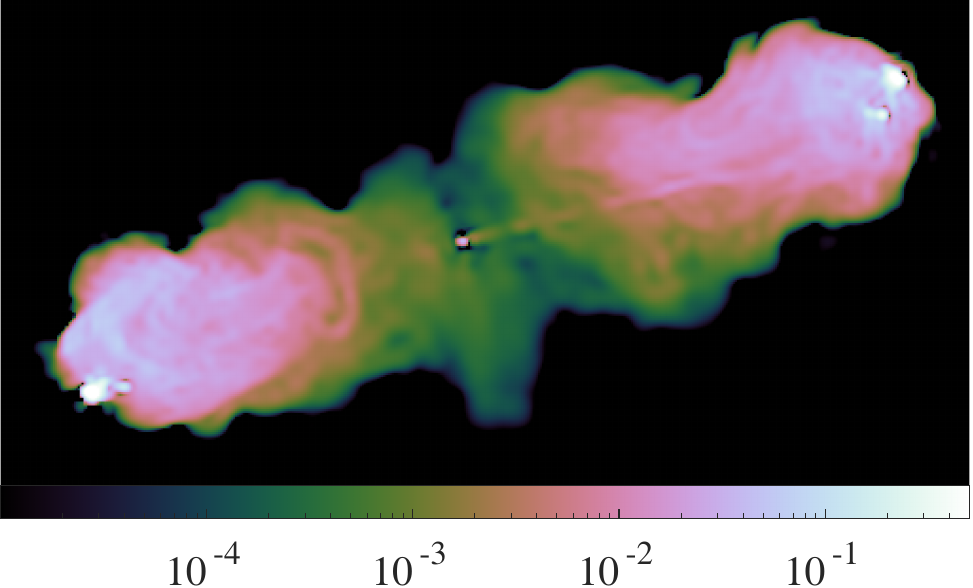}
            \raisebox{-0.45cm}{\includegraphics[width=0.49\textwidth,keepaspectratio]{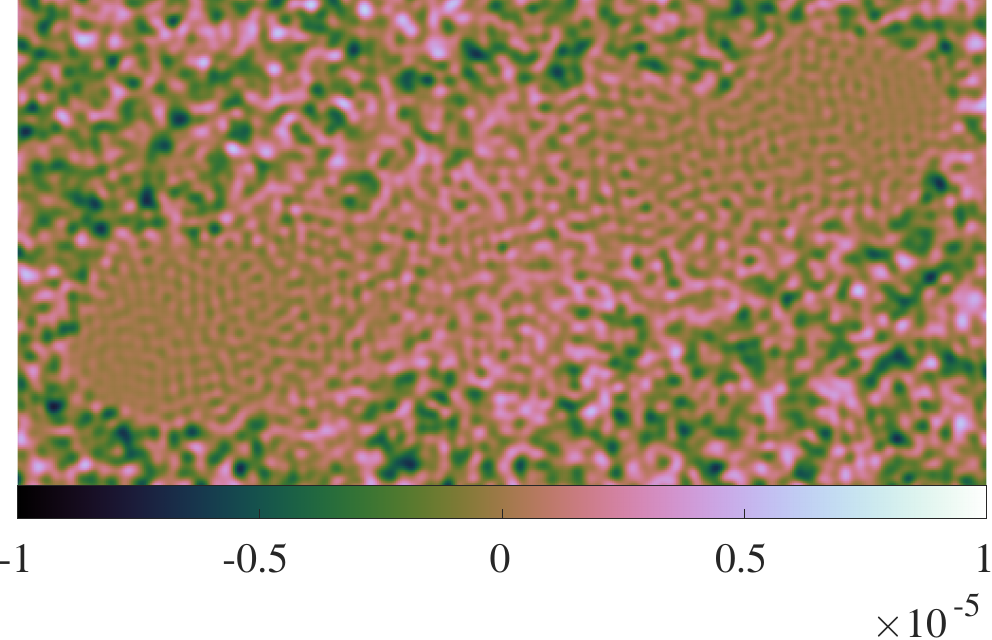}}\\
            \includegraphics[width=0.48\textwidth,keepaspectratio]{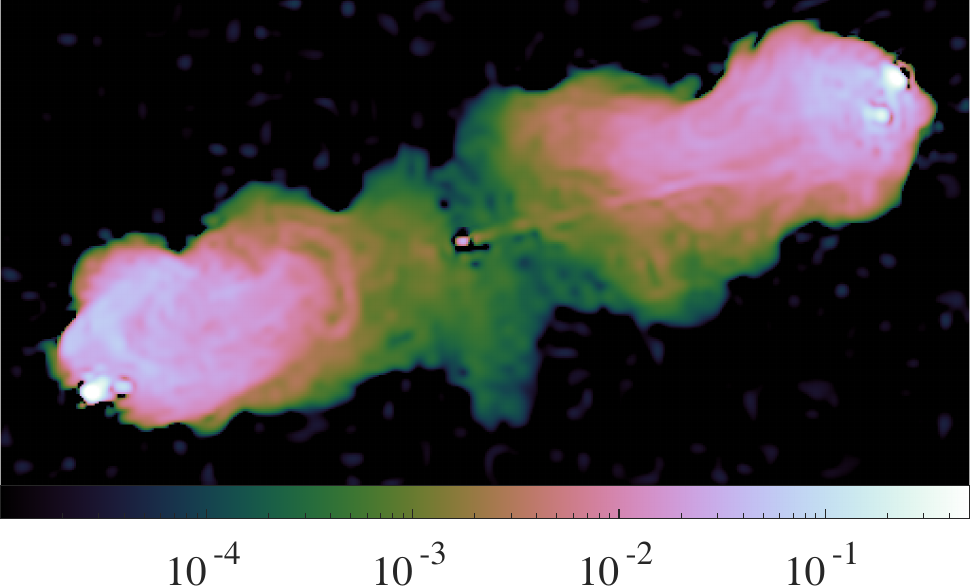}
            \raisebox{-0.45cm}{\includegraphics[width=0.49\textwidth,keepaspectratio]{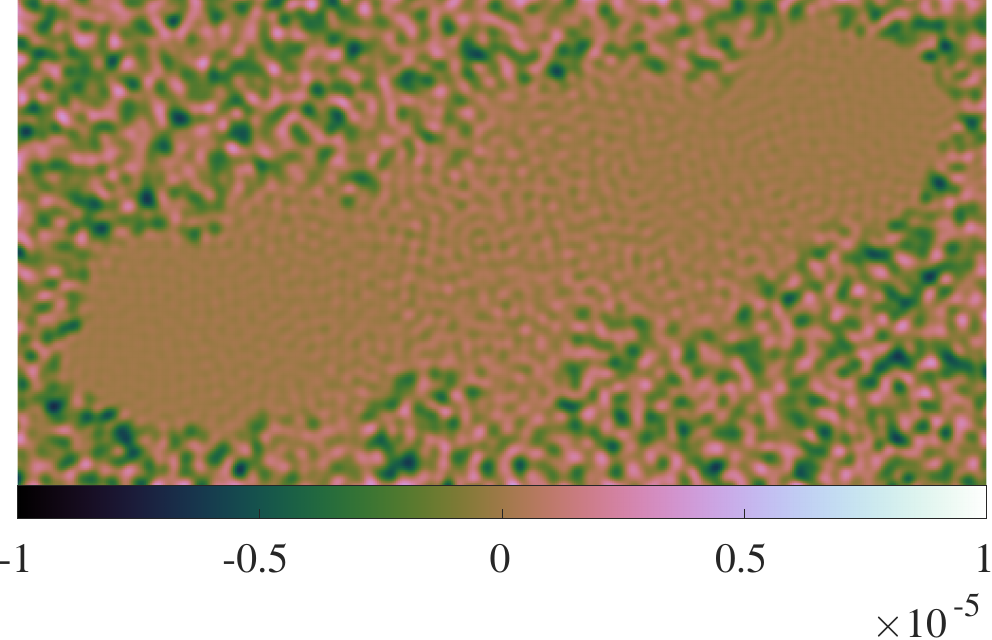}}\\
        \end{subfigure}
    \end{center}

    \vspace*{-0.2cm}

    \caption{Spectral faceting experiment: reconstructed images of channel $\nu_{1} = 2.052$ GHz of an image cube with $\nband=100$ spectral channels, obtained with Faceted HyperSARA (using $(\nfacet, \ncube)=(1,10)$), HyperSARA (\emph{i.e.} Faceted HyperSARA with $(\nfacet,\ncube)=(1,1)$) and SARA (\emph{i.e.} Faceted HyperSARA with $(\nfacet, \ncube, (\overline{\mu}_{c, q})_{c,q})=(1, \nband, \mathbf{0}_{\ncube \times \nfacet})$). Top row: ground truth image (Jy/pixel) displayed in $\log_{10}$ scale. In rows $2-4$ are reported the reconstructed images of Faceted HyperSARA, HyperSARA and SARA, respectively. From left to right, model image (Jy/pixel) displayed in $\log_{10}$ scale and residual image normalized by $\|\bs{\Phi}_{1}\|_{\text{S}}^2$, displayed in linear scale.}
    \label{fig:images_sub_1}
\end{figure*}

\begin{figure*}
    \begin{center}
        \begin{subfigure}{\textwidth}
            \includegraphics[width=0.48\textwidth,keepaspectratio]{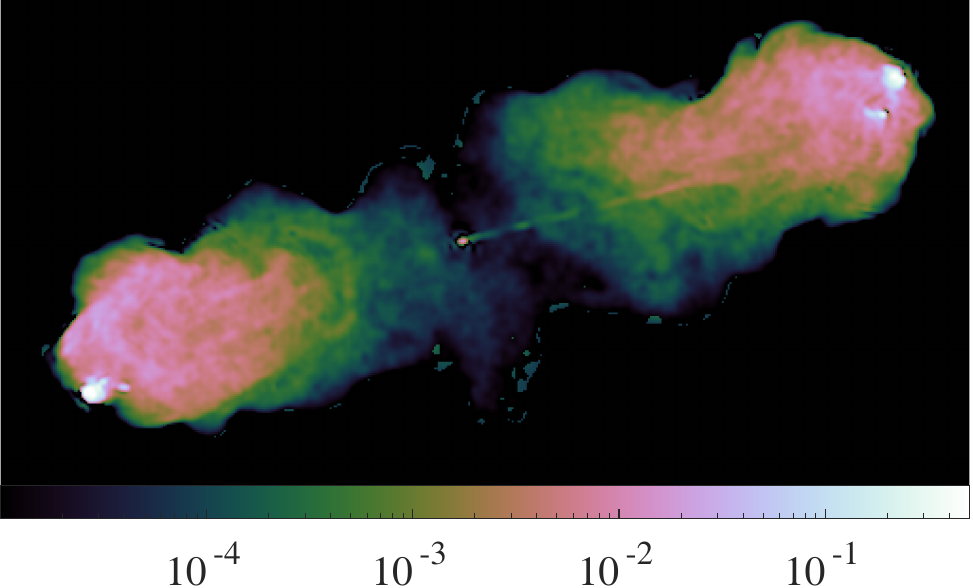}\\
            \includegraphics[width=0.48\textwidth,keepaspectratio]{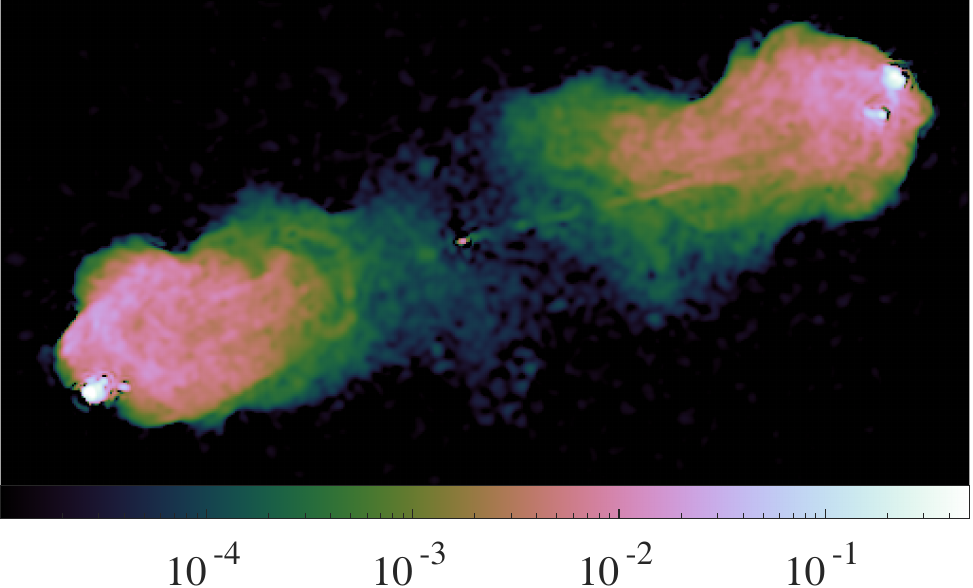}
            \raisebox{-0.45cm}{\includegraphics[width=0.49\textwidth,keepaspectratio]{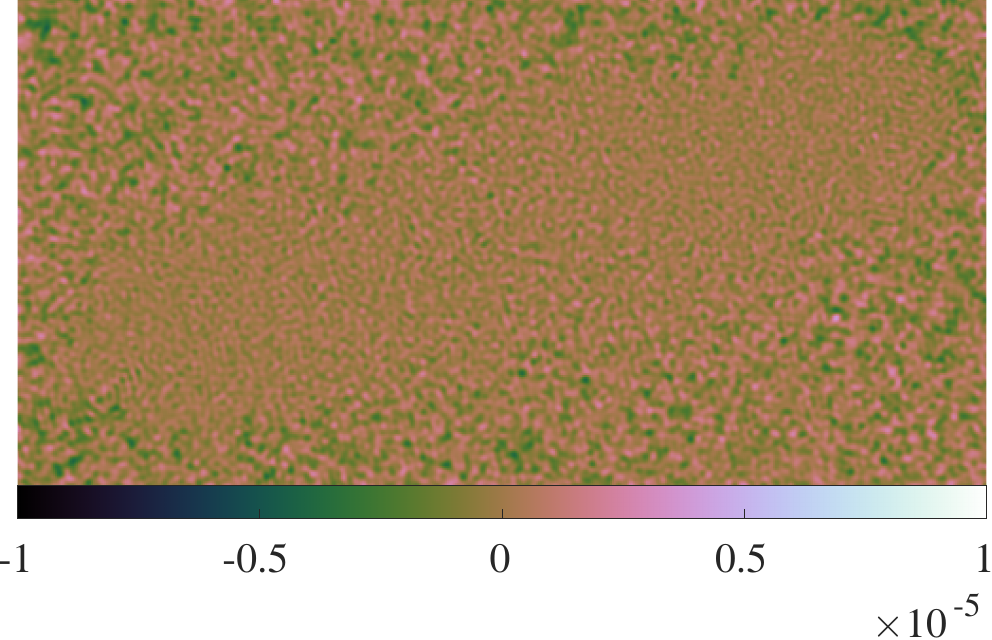}}\\
            \includegraphics[width=0.48\textwidth,keepaspectratio]{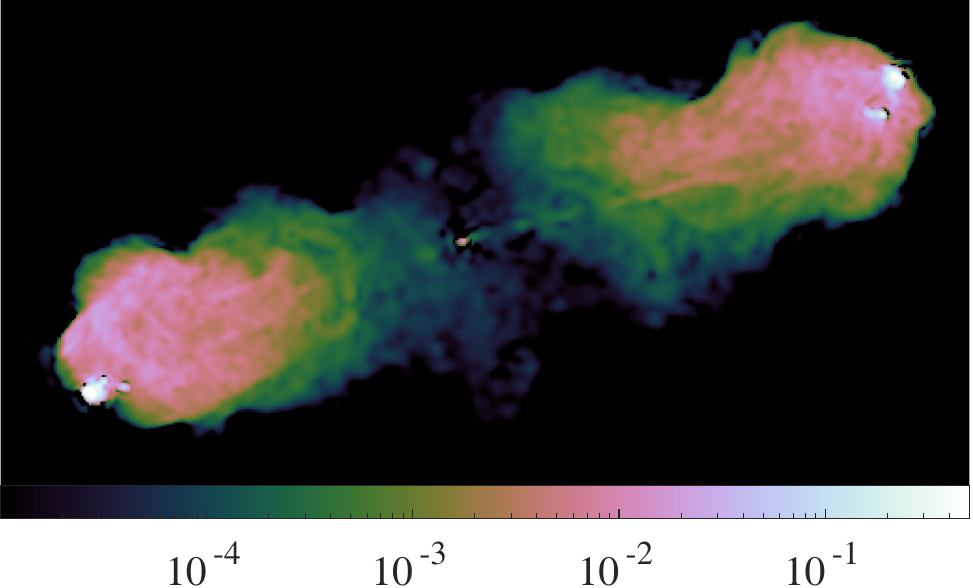}
            \raisebox{-0.45cm}{\includegraphics[width=0.49\textwidth,keepaspectratio]{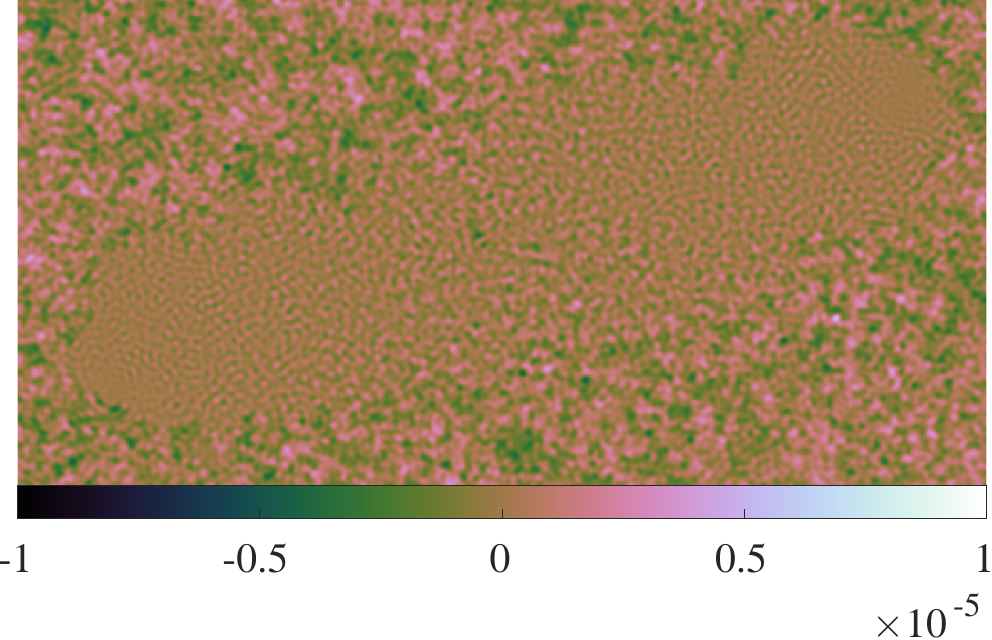}}\\
            \includegraphics[width=0.48\textwidth,keepaspectratio]{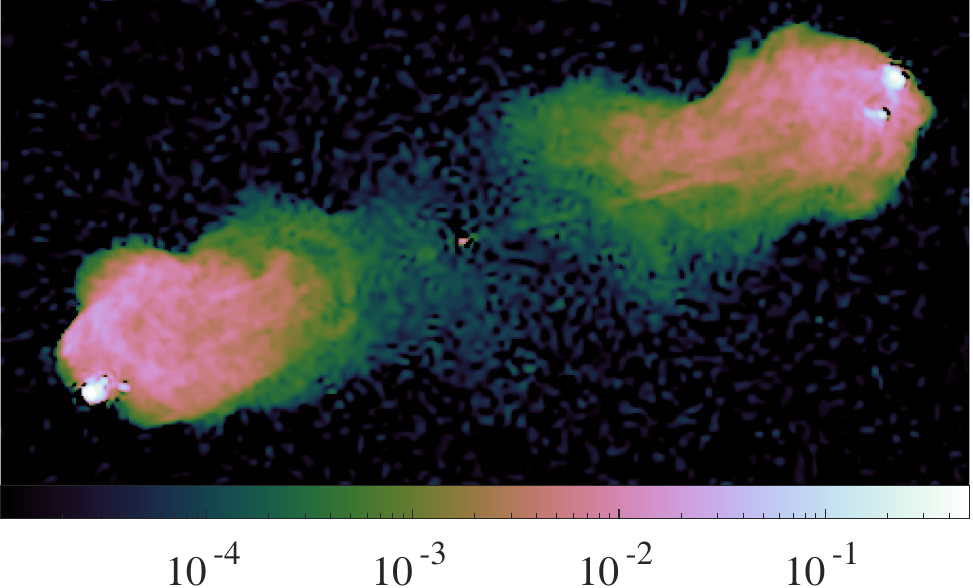}
            \raisebox{-0.45cm}{\includegraphics[width=0.49\textwidth,keepaspectratio]{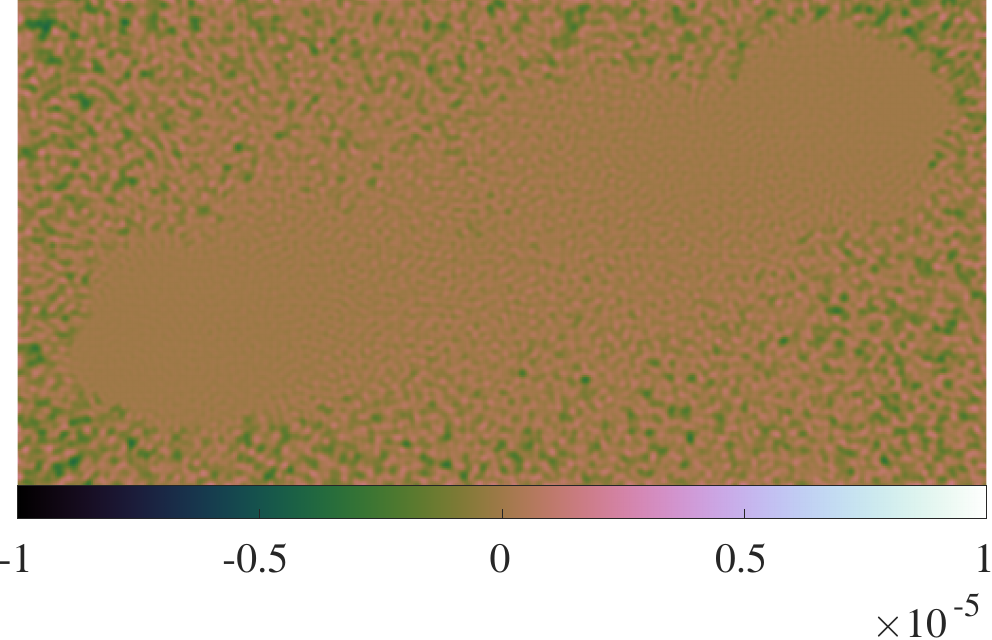}}
        \end{subfigure}
    \end{center}

    \vspace*{-0.2cm}

    \caption{Spectral faceting experiment: reconstructed images of channel $\nu_{100} = 3.636$ GHz of an image cube with $\nband=100$ spectral channels, obtained with Faceted HyperSARA (using $(\nfacet, \ncube)=(1,10)$), HyperSARA (\emph{i.e.} Faceted HyperSARA with $(\nfacet,\ncube)=(1,1)$) and SARA (\emph{i.e.} Faceted HyperSARA with $(\nfacet, \ncube, (\overline{\mu}_{c, q})_{c,q})=(1, \nband, \mathbf{0}_{\ncube \times \nfacet})$). Top row: ground truth image (Jy/pixel) displayed in $\log_{10}$ scale. In rows $2-4$ are reported the reconstructed images of Faceted HyperSARA, HyperSARA and SARA, respectively. From left to right, model image (Jy/pixel) displayed in $\log_{10}$ scale and residual image normalized by $\|\bs{\Phi}_{100}\|_{\text{S}}^2$, displayed in linear scale.}
    \label{fig:images_sub_2}
\end{figure*}

\begin{table*}
    \centering
    \resizebox{0.98\textwidth}{!}{%
        \begin{tabular}{lrrrrrrrr} \toprule
            & $\aSNR$ ($\pm \sSNR$) (dB)  & $\aSNR_{\log}$ ($\pm \sSNR_{\log}$) (dB) & CPU cores & PDFB iterations & $\runp$ (s) & $\run$ (h) & $\cpup$ (s) & $\cpu$ (h) \\ \midrule
            SARA
            & \textbf{19.76 ($\pm 3.19e+00$)} & $19.85 (\pm 1.80e+00$) & 1200 & 2205 & \textbf{0.55 ($\pm 4.59e-02$)} & 0.41 & \textbf{0.87 ($\pm 5.56e-02$)}  & 53.01 \\
            HyperSARA
            & \textbf{22.27 ($\pm 2.56e+00$)} & 23.57 ($\pm 1.29e-01$) & 16 & 3800 & \textbf{11.30 ($\pm 1.01e+00$)} & 12.01 & \textbf{64.71 ($\pm 2.42e+00$)} & 68.75 \\
            Faceted HyperSARA ($\ncube = 2$)
            & \textbf{21.77 ($\pm 2.51e+00$)} & 23.37 ($\pm 1.19e-00$) & 32 & 2400 & \textbf{5.68 ($\pm 4.49e-01$)} & 3.80 & \textbf{32.25 ($\pm 1.72e+00$)} & 43.18 \\
            Faceted HyperSARA ($\ncube = 5$)
            & \textbf{21.85 ($\pm 2.72e+00 $)} & 24.73 ($\pm 6.12e-01$) & 80 & 2380 & \textbf{2.67 ($\pm 4.35e-01$)} & 2.01 & \textbf{13.78 ($\pm 1.17e+00$)} & 45.74 \\
            Faceted HyperSARA ($\ncube = 10$)
            & \textbf{22.04 ($\pm 2.85e+00 $)} & 25.03 ($\pm 3.97e-01$) & 160 & 2540 & \textbf{1.53 ($\pm 2.90e-01$)} & 1.36 & \textbf{7.04 ($\pm 9.10e-01$)} & 49.58 \\
            \bottomrule
        \end{tabular}
    }
    \caption{Spectral faceting experiment: reconstruction performance of Faceted HyperSARA with a varying number of spectral sub-problems $\ncube$ and $\nfacet=1$, compared to HyperSARA (\emph{i.e.} Faceted HyperSARA with $(\nfacet,\ncube)=(1,1)$) and SARA (\emph{i.e.} Faceted HyperSARA with $(\nfacet, \ncube, (\overline{\mu}_{c, q})_{c,q})=(1, \nband, \mathbf{0}_{\ncube \times \nfacet})$). The results are reported in terms of
    average $\aSNR$ and $\aSNR_{\log}$ (both in dB with the associated standard deviation $\sSNR$ and $\sSNR_{\log}$), runtime and CPU time per iteration per sub-cube ($\runp$ and $\cpup$ in seconds, with the associated standard deviation over the iterations and the $C$ spectral subcubes). The number of CPU cores, the average number of PDFB iterations across all sub-problems until convergence and the total runtime and CPU time to reconstruct the full image cube are also reported. Note that the runtime and CPU time per iteration are reported as an average over all the sub-problems (since the solvers can be applied independently to distinct sub-problems). The total runtime is defined as the total amount of time needed to reconstruct the full image cube (\emph{i.e.}, the largest runtime taken by the compared methods over all the reconstructed sub-cubes). The reconstruction $\SNR$ and the runtime per iteration are highlighted in bold face.}
    \label{tab:spectral_faceting}
\end{table*}

%--------------------------------------------
\subsection{Results and discussion}

Performance assessment based on the results from Tables~\ref{tab:facet_overlap},~\ref{tab:facet_number} and~\ref{tab:spectral_faceting} is conducted in terms of reconstruction performance ($\SNR$ and $\SNR_{\log}$) and computing time. Since the following experiments rely on comparisons between the solution to different imaging problems, the total number of PDFB iterations until convergence (\emph{i.e.}, over all the reweighting steps) is reported to better interpret the total computing time required by each algorithm.

\subsubsection{Spatial faceting} \label{ssec:exp_spatial_faceting}

\begin{itemize}
    \item \textit{Varying spatial overlap:} The $\SNR$ and $\SNR_{\log}$ results reported in Table~\ref{tab:facet_overlap} show that spatial faceting gives a good reconstruction quality, that is much superior to SARA, and at the same level as HyperSARA, independently of the percentage of overlap. These results are also confirmed in both the reconstructed and residual images, reported in Jy/pixel in Figure~\ref{fig:images_overlap_1} and~\ref{fig:images_overlap_2} for the channels $\nu_1 = 2.052$~GHz and $\nu_\nband = 3.572$~GHz. In absence of overlap, and in this case only, facet artefacts emerge in the reconstructed image for some channels (see the areas delineated in white in Figure~\ref{fig:images_overlap_2}, where the horizontal and vertical marks at the panel edges identify the contour of each tile). This observation confirms the importance of the facet overlapping procedure, and its effectiveness in preventing faceting artefacts. Overall, an overlap of $10\%$ is already enough.
    From a computational point of view, Table~\ref{tab:facet_overlap} shows that the CPU time per iteration is higher for Faceted HyperSARA when compared to HyperSARA, yet the overall runtime is reduced by using more resources in parallel through faceting. In particular, the proposed faceting procedure allows for a more flexible use of the available resources compared to HyperSARA, in that the low-rankness prior no longer constrains variables of the size of the full image cube to be stored in and communicated from a single worker. The table also confirms that increasing the overlap size from $10\%$ to $50\%$ moderately affects the runtime per iteration and the CPU time per iteration. All the Faceted HyperSARA versions reported for this experiment require a comparable total number of PDFB iterations to converge.
    \item \textit{Varying number of facets $\nfacet$ along the spatial dimension:}
    The reconstruction performance and computing time reported in Table~\ref{tab:facet_number} show that Faceted HyperSARA gives a stable reconstruction performance independently of the number of facets. The runtime per iteration decreases significantly when the number of facets increases, confirming the efficiency of the parallelization enabled by the faceting approach. Although the total number of PDFB iterations increases with the number of facets, the significant reduction in the per iteration runtime leads to an overall reduction of the total runtime. This paves the way to a scalable approach where the number of spatial facets could be increased proportionally to the spatial dimension of the image, possibly up to hundreds or thousands, with the reconstruction parallelized over as many cores of a large computing system.
\end{itemize}

\subsubsection{Spectral faceting}
The $\SNR$ and $\SNR_{\log}$ results reported in Table~\ref{tab:spectral_faceting} show that Faceted HyperSARA using channel-interleaved facets retains the overall reconstruction performance of HyperSARA. This is also qualitatively confirmed by the reconstructed and residual images reported in Figures~\ref{fig:images_sub_1} and~\ref{fig:images_sub_2} (in Jy/pixel). While the $\aSNR$ value and residual images deteriorate slightly when interleaving is applied, the $\aSNR_{\log}$ values are actually higher in some cases. This suggests that the hyperspectral prior is effective at compensating for the loss of sensitivity when data are divided up between sub-cubes. The reconstruction quality of SARA is at no surprise significantly lower due to a prior failing to leverage any inter-channel correlation to build sensitivity. 
Spectral faceting thus appears computationally attractive, in that it preserves the overall imaging quality up to an already significant amount of interleaving (see discussion in Section~\ref{sec:spectral_faceting}). This paves the way to a scalable approach (complementary to the one offered by spatial faceting) where the number of spectral facets would be increased proportionally to the spectral dimension of the image cube, again possibly up to hundreds or thousands, enabling to divide up the reconstruction process over as many independent optimization problems.

Note that an alternative (and in fact complementary) approach to spectral faceting when targeting sources with smooth spectra slowly varying across the frequency band of observation is firstly, to collapse the channels into a severely reduced number of effective channels (thus targeting an image cube with possibility tens of channels rather than tens of thousands), and secondly, to leverage data dimensionality reduction techniques to control the data size per channel \citep{Kartik2017}.

%% ==============================================================
%% CONCLUSION AND FUTURE WORK
%% ==============================================================
\section{Conclusion} \label{sec:conclusion}

We have proposed the Faceted HyperSARA approach based on convex optimization, which leverages a spatio-spectral facet prior model for wide-band intensity image reconstruction in radio interferometry. The underlying regularization encodes a facet-specific prior model to ensure high quality of the reconstructed images, while allowing the bottleneck induced by the size of the image cube to be efficiently addressed via parallelization. Experiments conducted on synthetic data and image cubes close to Gigabyte size confirm that faceting provides a significant improvement in the parallelization capability compared to the original HyperSARA algorithm~\citep{Abdulaziz2019}. The results also confirm the relevance of the heuristic strategy developed to enable  an automatic setting of the regularization parameters. This paves the way towards an image formation method that (i) can scale to large image dimensions, and (ii) does not require user expertise, or intervention, to set key regularization parameters.

Further work is required to confirm the practical scalability of Faceted HyperSARA, in particular when accounting for inter-node communication requirements at scales where the algorithm needs to be parallelized across multiple compute nodes for each independent sub-problem. At the extreme scales expected in the SKA era, further accelerations of the RI measurement operators should also be investigated, ranging from applying data dimensionality reduction techniques to control the data size per channel \citep{Kartik2017}, to breaking the remaining bottleneck of the image channel-size Fourier transforms.

In a following companion paper entitled ``Parallel faceted imaging in radio interferometry via proximal splitting (Faceted HyperSARA): II. Real data proof-of-concept and code'', new functionalities of Faceted HyperSARA will be introduced, which are key to its use on large real datasets. These include the ability to account for DDE kernels in the image formation process, and to work with gridded visibilities for data dimensionality reduction. Faceted HyperSARA will be showcased in various settings to reconstruct a large image cube from VLA data, in comparison with both SARA and CLEAN. Importantly, a fully documented MATLAB toolbox of faceted HyperSARA (including HyperSARA, and SARA as limit cases) will also be introduced, as part of a wider effort towards the adoption of the algorithms of the SARA family by the community.

\section*{Acknowledgements}
This work was supported by the Engineering and Physical Sciences Research Council (EPSRC, grants EP/M011089/1, EP/M008843/1, EP/M019306/1, EP/T028270/1) and the Swiss-South Africa Joint Research Program (IZLSZ2\_170863/1). The computing resources are provided by the Cirrus UK National Tier-2 HPC Service at the Edinburgh Parallel Computing Centre (EPCC, http://www.cirrus.ac.uk) funded by the University of Edinburgh and EPSRC (EP/P020267/1), with time allocation under the SUSA project. The authors warmly thank Ming Jiang, Jean-Philippe Thiran, and Adrian Jackson for the insightful discussions.

\section*{Data availability}
All data and code to reproduce the experiments reported in both this and the companion paper are available with the MATLAB toolbox from the Puri-Psi webpage~\citep{puripsi}.
%%%%%%%%%%%%%%%%%%%%%%%%%%%%%%%%%%%%%%%%%%%%%%%%%%

%%%%%%%%%%%%%%%%%%%% REFERENCES %%%%%%%%%%%%%%%%%%
\bibliographystyle{mnras}
\bibliography{strings_all_ref,all_ref}

%%%%%%%%%%%%%%%%%%%% APPENDIX %%%%%%%%%%%%%%%%%%%%
% \appendix
% ...

%%%%%%%%%%%%%%%%%%%%%%%%%%%%%%%%%%%%%%%%%%%%%%%%%%

% Don't change these lines
\bsp	% typesetting comment
\label{lastpage}
\end{document}